\documentclass[a4paper,12pt]{article}
\pdfoutput=1 
\usepackage{jheppub}
\usepackage{amssymb}
\usepackage{amsfonts}
\usepackage{amsmath}
\usepackage{color}
\usepackage{ulem}
\usepackage[english]{babel}
\usepackage[T1]{fontenc}
\usepackage[utf8]{inputenc}
\usepackage{float}
\usepackage[caption = false]{subfig}

\newcommand{\ksymbol}{\kappa} 
\newcommand{\scsymbol}{\alpha_\lambda}
\newcommand{\kscsymbol}{\alpha_\ksymbol}
\newcommand{\wscsymbol}{\alpha_w}
\newcommand{\VgIRsymbol}{V_\mathrm{IR}}
\newcommand{\WIRsymbol}{W_\mathrm{IR}}
\newcommand{\kIRsymbol}{\bar \ksymbol_0}
\newcommand{\kUpsymbol}{\ksymbol_1}
\newcommand{\wUpsymbol}{w_1}
\newcommand{\kIRpsymbol}{\bar \ksymbol_1}
\newcommand{\wIRsymbol}{\bar w_0}
\newcommand{\wIRpsymbol}{\bar w_1}
\newcommand{\lambdapar}{W_\tau}

\title{Regge theory in a Holographic dual of QCD in the Veneziano Limit}
\author[a]{Artur Amorim,\!} 
\author[a]{ Miguel S. Costa,\!} 
\author[b,c]{ Matti J\"arvinen\,} 
\affiliation[a]{Centro de F\'{\i}sica do Porto e Departamento de F\'{\i}sica e Astronomia da Faculdade de Ci\^encias da Universidade do Porto, 
Rua do Campo Alegre 687, 4169-007 Porto, Portugal} 
\affiliation[b]{Asia Pacific Center for Theoretical Physics, Pohang 37673, Republic of Korea} 
\affiliation[c]{Department of Physics, Pohang University of Science and Technology, Pohang 37673, Republic of Korea}
\date{}

\preprint{APCTP Pre2021 - 002}

\abstract{We initiate the study of Regge theory in a bottom-up holographic model 
for 
QCD in the Veneziano limit, where the backreaction of the quarks to the gluon dynamics is included. 
We determine the parameters of the model by carrying out a precise fit to the meson spectrum in QCD. 
The spectrum for spin-one and pseudoscalar mesons is well reproduced. 
We then generalise the model to incluce 
higher spin fields in the bulk trajectories dual to the Pomeron and meson Regge trajectories 
at 
the boundary. With this setting, we 
fit the masses of the mesons with 
spins $J=2$, $3$, and $4$,
as well as
the experimental data of the total cross-sections $\sigma(\gamma \gamma \to X)$, $\sigma(\gamma p \to X)$ and $\sigma(p p \to X)$. 
For the cross sections we  obtain 
a $\chi^2_{\mathrm{d.o.f.}}$ of $0.74$ for a total of 199 experimental points.}

\begin{document}
\maketitle

\section{Introduction}

Since it was conjectured that the QCD Pomeron is dual to the graviton Regge trajectory~\cite{Brower:2006ea}, holographic techniques have been successfully applied to the description of QCD processes where Pomeron exchange dominates~\cite{BallonBayona:2007qr, Hatta:2007he, Cornalba:2008sp, Pire:2008zf,  Albacete:2008ze, Hatta:2008st, Brower:2008ix, Levin:2009vj, Brower:2009bh, Gao:2009ze, Hatta:2009ra, Kovchegov:2009yj, Avsar:2009xf, Domokos:2009hm, Cornalba:2009ax, Dominguez:2009cm, Cornalba:2010vk, Betemps:2010ij, Gao:2010qk, Kovchegov:2010uk, Levin:2010gc, Domokos:2010ma, Brower:2010wf, Costa:2012fw, Brower:2012mk, Stoffers:2012zw, Costa:2013uia, Anderson:2014jia, Koile:2014vca, Koile:2015qsa, Ballon-Bayona:2015wra, Kovensky:2016ryy, Ballon-Bayona:2017vlm, Nally:2017nsp, Kovensky:2018xxa, Lee:2018zud, Amorim:2018yod, Kovensky:2018gif, Mamo:2019mka, FolcoCapossoli:2020pks, Amorim:2021ffr}. Once the external scattering states and the dynamics of the higher spin $J$ fields of the graviton trajectory are modelled, comparisons can be made with experiment, provided we are in a kinematical window where QCD is dominated by a gluon rich medium. In this regime the Bjorken variable $x$ is small, or the Mandelstam variable $s$ is large,  corresponding to high center of mass energies. 
 
In order to describe the total cross-section data of hadronic processes in QCD, one also includes, besides the Pomeron trajectory, a meson trajectory (see, for e.g, \cite{Donnachie:2002en}). This trajectory can be obtained by a linear fit of the meson spins against the meson squared masses. With the resulting straight line, one extrapolates from $t > 0$ to $t \leq 0$ in order to make predictions in the scattering region. In particular, for the total cross-section we are interested in the value of the trajectory for $t = 0$, also known as the meson intercept, and the above linear fit yields an intercept of $0.55$. There is no theoretical justification to assume that the meson trajectory is linear aside from the fact that it has described successfully scattering data, provided $t$ is not too negative. In this work we  
use holography to study this issue by first fitting the meson spectrum for $J =0,1,2,3,4$ and check if the resulting holographic intercept is able to describe the experimental data of the total cross-sections of $\gamma \gamma$, $\gamma p$ and $pp$ scattering. In previous works~\cite{Ballon-Bayona:2017vlm, Amorim:2018yod, Amorim:2021ffr} we have studied the dynamics of the higher spin $J$ fields in the graviton Regge trajectory by generalising the bulk graviton equation of motion. This was done using effective field theory inspired by Regge theory of a $5D$ string theory. In this work we will not only follow the same procedure for the Pomeron trajectory, but also apply it to the meson trajectory by generalising to higher spin the equation of motion of the bulk field dual to the vector mesons.

Before we start such procedure, we need to guarantee that our model is describing with accuracy the spectrum of the vector mesons. This will be done by considering 
the extension of the Improved Holographic QCD model of~\cite{Gursoy:2007cb, Gursoy:2007er}, 
with a backreacted quark sector~\cite{Bigazzi:2005md,Casero:2007ae},  as
presented in \cite{Jarvinen:2011qe}. This model 
(V-QCD)  consists of 
five-dimensional dilaton gravity dilaton coupled to a tachyon described in terms of a generalised Sen-like tachyonic Dirac-Born-Infeld (DBI) action~\cite{Sen:2004nf}.  
Our numerical solution includes the full backreaction of the tachyon in the dilaton and metric.
The asymptotic behavior of the model at weak and strong coupling is chosen such that various generic features of QCD, such as asymptotic freedom and confinement, are reproduced~\cite{Gursoy:2007er,Jarvinen:2011qe}. 
The remaining 
free parameters 
of the model will be determined 
through an extensive comparison of  
the spectrum of the quadratic fluctuations against the experimental meson masses.

This paper is organized as follows. In section~\ref{seq:hvqcd_sum} we discuss in detail the holographic model,  
as well as 
how to compute the spectrum of the quadratic fluctuations. This section ends with a fit to the meson spectrum, fixing our background fields for the remaining of the paper. In section~\ref{seq:hol_xsections} we derive holographic expressions for the total cross-sections that will be used later to fit data from the Particle Data Group~\cite{Zyla:2020zbs}. In section~\ref{seq:spin_j_dynamics} we focus on the  holographic duals of the pomeron and meson trajectories, and in particular in constructing the analytic continuation of the spin $J$ equations that govern the dynamics of fields in these trajectories. These equations contain two parameters that will be fixed by the soft-pomeron intercept and the spin $J=2,3,4$ meson masses in section~\ref{seq:xseq_fits}. This fixes the pomeron and meson kernels that are used in the total cross-section fits. We discuss our results and suggest further work in section~\ref{seq:conclusions}.

\section{Holographic model for QCD in the Veneziano Limit}
\label{seq:hvqcd_sum}

We consider a slight generalisation of Quantum Chromodynamics
which consist  of a gauge field  in the adjoint representation of $SU(N_c)$  
coupled to $N_f$ fermions (quarks) in the fundamental representation of $SU(N_c)$. 
This generalisation has been studied in great depth in the 't Hooft large-$N_c$ limit,  where $N_c \rightarrow \infty$ and $\lambda = g_{\mathrm{YM}}^2 N_c$ and $N_f$ are kept fixed. This limit is also known as the  \textit{quenched} limit since nontrivial quark contributions 
to observables
are suppressed in powers of $\frac{N_f}{N_c} \rightarrow 0$.
Another interesting large-$N_c$ limit is 
the Veneziano limit~\cite{Veneziano:1979ec}  
where
\begin{align}
N_c \rightarrow \infty\, , \quad N_f \rightarrow \infty \, , \quad \frac{N_f}{N_c} = x  \,, \quad \lambda = g_{\mathrm{YM}}^2 N_c \, .
\end{align} 
with $x$  and $\lambda$ fixed. 
In this limit the quark contributions are not suppressed, and their backreaction to the gluon dynamics must be taken into account. 

A holographic dual  (V-QCD)  
that reproduces 
several expected features of QCD in the Veneziano limit
was presented in \cite{Jarvinen:2011qe}. It consists of a system of a dilaton and tachyon coupled to five-dimensional gravity. Let us  first discuss the field content and the action of the model, and then present the precise structure of the various potentials appearing in the action.

\subsection{The model}

The action in the gravitation sector is the same as in the Improved Holographic QCD (IHQCD) model  
\cite{Gursoy:2007cb,Gursoy:2007er}.  
We work at zero temperature so that Poincar\'e invariance is intact. The  metric Ansatz is therefore
\begin{equation}
\label{eq: metric definition}
ds^2 = g_{ab} dx^a dx^b = e^{2 A(z)} \left( \eta_{\mu \nu} dx^\mu dx^\nu + d z^2  \right),
\end{equation}
where the warp factor $A$ is identified with the logarithm of the energy scale in the field theory at the boundary. 
The exponential of the dilaton field $\lambda = e^\Phi$ is dual to the ${\mathrm{Tr}} F^2$ operator
with its background value equal to the 't Hooft coupling (near the boundary where the coupling can be unambigously defined).
The action for the metric and dilaton fields is given by  five-dimensional Einstein gravity coupled to a scalar field,
\begin{equation}
S_g =  M_p^3 N_c^2 \int d^5x \sqrt{-\det g} \left[ R - \frac{4}{3} \frac{{\left( \partial \lambda \right)}^2}{\lambda^2}+ V_g ( \lambda) \right]\, ,
\label{eq:sg_action}
\end{equation}
where $M_p$ is the five-dimensional Planck scale. 
The dilaton potential $V_g(\lambda)$ will be specified below. We choose the potential to be qualitatively similar to that studied in~\cite{Gursoy:2008bu,Gursoy:2008za} -- see~\cite{Gubser:2008ny,DeWolfe:2010he} for an alternative approach focused on the fit to QCD thermodynamics.

To add matter we insert space-filling $D_4 - \bar{D_4}$ branes that give rise to a tachyon field $T$
and the gauge fields $A_L$, $A_R$ living on the branes~\cite{Bigazzi:2005md,Casero:2007ae}.  
A similar approach has been considered in the probe limit in \cite{Iatrakis:2010zf,Iatrakis:2010jb}, 
and also in the  
Witten-Sakai-Sugimoto model \cite{Bergman:2007pm,Dhar:2007bz,Dhar:2008um,Jokela:2009tk}. In the boundary theory the operators with lowest dimension involving fermions are $\bar{\psi}^{i}_R \psi^j_L$ with spin 0 and the two spin 1 conserved currents 
$\bar{\psi}^i_L \gamma^\mu \psi^j_L$ and $\bar{\psi}^i_R \gamma^\mu \psi^j_R$, 
where $i$, $j$ are the flavour indices.  
The spin 0 and spin 1 operators  
are dual to bulk complex scalars $T_{ij}$ and two bulk gauge fields $A^{\mu}_{L ,\, ij}$ and $A^{\mu}_{R,\, ij}$, 
respectively.  
The tachyon  
transforms as $(N_f, \bar{N}_f)$ of the flavour symmetry $U\left(N_f\right)_R \times U \left(N_f\right)_L$, while the fields $A^\mu_{L, i j}$ and $A^\mu_{R, i j}$ transform in the adjoint representations of $U\left(N_f\right)_L$ and  $U \left(N_f\right)_R$, respectively. In string theory the three bulk fields can be modelled by considering $N_f$ flavour branes ($R$) and $N_f$ flavour antibranes ($L$).  In this configuration the complex scalar fields $T_{ij}$ are the lowest modes of open strings with one end in a D-brane and another in the anti-D-brane, while the bulk gauge  fields are the lowest open string modes with both ends on a D-brane or on an anti-D-brane. 
The  system obeys a tachyonic Dirac-Born-Infeld (DBI) action~\cite{Casero:2007ae,Jarvinen:2011qe,Arean:2013tja}
 \begin{align}
\label{eq: DBI action}
S_\mathrm{DBI} =&  - 
\frac{M_p^3 N_c}{2}  
\!\int\! d^5 x \, \mathbf{Str}\! \left[ V_f\! \left(\lambda, T^{\dagger} T\right)\! \sqrt{-\det \Big( g_{ab} + \ksymbol(\lambda)D_{(a}T^{\dagger}D_{b)}T + w(\lambda) F^{L}_{ab}\Big)} \right. \notag \\ 
&\left. + V_f \big(\lambda, T T^{\dagger}\big) \sqrt{-\det\Big( g_{ab} + \ksymbol(\lambda)D_{(a}TD_{b)}T^{\dagger} + w(\lambda) F^{R}_{ab}\Big)}  \, \right] ,
\end{align}
where $\mathbf{Str}$ is the symmetric trace over the 
(hidden) flavour indices  
and the determinant is taken with respect to 
the five dimensional space-time indices $a$, $b$  
(since we are going to work up to quadratic order we can 
actually replace the symmetric trace by the usual trace of matrices for the purposes of this article). 
The functions $V_f$, $\kappa$, and $w$ will be given explicitly below.
The normalisation convention for the symmetrisation of indices is $F_{(a}G_{b)} = \frac{1}{2}(F_{a}G_{b}+F_{b}G_{a})$. The covariant derivative terms are given by
\begin{equation}
D_a T = \partial_a T - i T A^L_a + i A^R_a T \, , \qquad   D_a T^{\dagger} = \partial_a T^{\dagger} - i  A^L_a T^{\dagger}+ i T^{\dagger} A^R_a \,,
\end{equation}
and the field strengths by
\begin{align}
F^{L,R} = d A^{L,R} - i A^{L,R} \wedge A^{L,R}\,.
\end{align}
In this work we are assuming that the light quark masses are equal and under this assumption the tachyon is just $T = \tau \, \mathbb{I}_{N_f}$. Furthermore,
 for the QCD vacuum we have  $A^R_a = 0 = A^L_a$. Using these conditions in the action (\ref{eq: DBI action}) we obtain the flavour action
\begin{equation}
S_f = - x M_p^3 N_c^2 \int d^5 x\, V_f(\lambda, \tau) \sqrt{-\det\Big(g_{ab} + \ksymbol(\lambda) \partial_a \tau \partial_b \tau \Big)}\,.
\label{eq:sf_equation}
\end{equation}
The brane action also contains a Wess-Zumino term~\cite{Casero:2007ae} which we will not need  as it does not contribute to the background solutions or mass spectra considered here.

There is also an additional pseudo-scalar axion field $a$, which is dual to the operator ${\mathrm{Tr}} F\wedge F$  and therefore sources the $\theta$ angle in QCD~\cite{Casero:2007ae}. Its action takes the form~\cite{Gursoy:2007cb,Gursoy:2007er,Arean:2013tja,Arean:2016hcs}
\begin{equation}
\label{eq:axion_action}
S_a = - \frac{M_p^3 N_c^2}{2} 
\int d^5 x \sqrt{-\det g}\, Z(\lambda)\! \left[ \partial_b a - x \left(\!V^{(a)}\!(\lambda, \tau)( \hat A^L_b\!-\!\hat A^R_b) - \rho \,\partial_b V^{(a)}\! (\lambda, \tau) \right)\!\right]^2 ,
\end{equation}
where $\hat A_b^{L,R}$ are the singlet fields, $\hat A_b^{L,R} = \mathrm{Tr}A_b^{L,R}/N_f$, 
the potential $V^{(a)}\! (\lambda, \tau)$ will be specified below,
and
we allowed the  tachyon to have an overall phase $\rho$, i.e. we took $T = \tau e^{i \rho} \mathbb{I}_{N_f}$.

\subsection{Choice of potentials} 

Let us now discuss the choices for the various potentials ($V_g$, $V_f$, $\ksymbol$, $w$, $V^{(a)}$ and $Z$). 
The generic picture is that the leading IR asymptotics of the various functions is chosen to agree with known features of QCD (which will be soon specified). The UV asymptotics is set by rough agreement with perturbative QCD, in particular by the perturbative UV dimensions of the QCD operators. For intermediate scales  the functions need to be determined by and extensive comparison to experimental and lattice QCD data; in this article  
the functions will be fitted to   
the meson spectrum in QCD. In this section we will present the Ans\"atze for these functions which will obey the asymptotics at small and large $\lambda$ determined by qualitative comparison to QCD, and the fit to data will be carried out in section~\ref{sec:fit}.

We start with the action for the gluon sector, i.e. the action of improved holographic QCD.
In IHQCD the dynamics of the dilaton is set by its potential $V_g(\lambda)$, which is constrained \cite{Gursoy:2007cb,Gursoy:2007er}
in the UV to reproduce the YM $\beta$-function and in the IR to 
yield  confinement and 
a ``good'' singularity in the classification of~\cite{Gubser:2000nd}.  
In this work we will take \cite{Alho:2015zua,Jokela:2018ers}
\begin{equation}
V_g(\lambda) = 12 + V_1 \, \lambda + V_2 \, \frac{\lambda^2}{1 + \frac{\scsymbol \, \lambda}{\lambda_0}} + 3 \VgIRsymbol \, e^{-\frac{\lambda_0}{\scsymbol \, \lambda}}  \frac{\lambda^{4/3}}{4 \pi^{8/3}}\sqrt{\log\left(1+\frac{\scsymbol \, \lambda}{\lambda_0}\right)} \, ,
\end{equation}
where
\begin{equation}
 V_1 = \frac{44}{9 \pi^2} \, ,\qquad  V_2 = \frac{4619}{3888 \pi^4} \, , \qquad \lambda_0 = 8 \pi^2 \, .
\end{equation}
The values of $V_1$ and $V_2$ are fixed by the gluon sector contribution to the QCD $\beta$-function, while the parameters $\scsymbol$ and $\VgIRsymbol$ will be fitted by the spectrum, in particular by computing meson  mass ratios and comparing them to experimental results, following \cite{Gursoy:2009jd}.  
The spectrum of the theory can be 
found  
by analysing the action of the quadratic fluctuations around the background solution followed by the reduction to the four-dimensional dynamics. For example, in the case of pure Yang-Mills (i.e. $x = 0$), the quadratic fluctuations are dual to glueballs with quantum numbers $J^{PC} = 0^{++}, 2^{++}$ and with $J^{PC} = 0^{-+}$ by considering an axion term. For pure Yang-Mills  we have found that $\scsymbol \approx 2.504$  
and   $\VgIRsymbol \approx 3.478$  
reproduce\footnote{We will instead determine these parameters by a global fit to the meson spectrum as we will explain in section~\ref{sec:fit}.} the lattice ratios of 1.46 and 1.87 of $m_{2^{++}} / m_{0^{++}}$ and $m_{0^{*++}} / m_{0^{++}}$ respectively. In the IR,
$V_g \sim \lambda^{\frac{4}{3}} (\log \lambda)^{\frac{1}{2}}$, which gives  linear asymptotic trajectories for glueballs.

The DBI action that we  described above is analogous to the flat space Sen action for the $D - \bar{D}$ system~\cite{Sen:2004nf}. Since we are in the presence of a curved space-time  and other non-trivial background fields 
which fully backreact to the metric,  
we correct it by including the general potentials $V_f(\lambda, \tau)$, $\ksymbol (\lambda)$ and $w(\lambda)$. However these potentials must satisfy some properties. The tachyon potential $V_f$ is expected to have a regular series expansion in $\lambda$ and $\tau$ near the boundary (i.e. $\lambda \rightarrow 0, \, \tau \rightarrow 0$)~\cite{Jarvinen:2011qe}
\begin{equation}
V_f(\lambda,\tau) = V_0 (\lambda) + V_1 (\lambda) \tau^2 + \mathcal{O}(\tau^4)\,,
\end{equation} 
and to vanish exponentially in the IR when $\tau \rightarrow \infty$~\cite{Arean:2016hcs}. In particular in the flat space string theory $V_s \sim \frac{1}{\lambda} e^{- \mu \tau^2}$. Our Ansatz for $V_f$ is
\begin{equation}
 V_f(\lambda, \tau) = V_{f0} (\lambda) \,V_\tau (\tau)\,,
 \end{equation}
 where 
\begin{align}
\quad V_\tau (\tau) &= \left( 1 + a_1 \tau^2\right) \,e^{- a_2 \tau^2}\,, 
\label{eq:V_tau}
\\ \notag
V_{f0}(\lambda)& = W_0 + W_1 \lambda + W_2\, \frac{\lambda^2}{1+ \frac{\scsymbol \, \lambda}{\lambda_0}} + \frac{3 W_{IR}}{16 \pi^4} {\left(\scsymbol \, \lambda \right)}^2 e^{- \frac{\lambda_0}{\scsymbol \, \lambda}} \left( 1 + \frac{\lambda_0 W_1}{\scsymbol \, \lambda}\right) 
\end{align}
with
\begin{equation}
W_1 = \frac{24 + \left(11- 2 x\right) W_0}{27 \pi^2}\,, 
\quad 
W_2 =  \frac{24 (857 - 46 x) + W_0 (4619 - 1714 x + 92 x^2)}{46656 \pi^4}\,.
 \end{equation}
This Ansatz is identical to that considered in~\cite{Jokela:2018ers} except for the introduction of the coefficients $a_i$ in $V_\tau(\tau)$. These are motivated by the observation that the correct mass gap of the mesons at large quark mass can only be reproduced if the coefficient of $\tau^2$ in the exponent determining the large-$\tau$ asymptotics of $V_\tau$ (i.e. $a_2$ above) differs from the second order series coefficient of $V_\tau$ at small $\tau$ (here $a_2-a_1$)~\cite{Jarvinen:2015ofa}. Notice that one of these coefficients can be 
eliminated 
by adjusting the normalisation of the $\tau$ field.

As both $\ksymbol (\lambda)$ and $w(\lambda)$ are coupling functions under the square root of the DBI action we expect them to have similar qualitative behaviour. On the other hand, in order to have the correct UV dimension of the $\bar{q}q$ operator we need to impose
\begin{equation}
\ksymbol (0)  =8 \,\frac{a_2 - a_1}{ 12 - x W_0 }\,.
\end{equation}
The IR asymptotics of the potentials $\ksymbol$ and $w$ directly affect the meson spectrum. We are interested in the case where the mesons have an asymptotic linear spectrum and  the meson towers have the same asymptotics. This can be achieved with the following IR asymptotics $\ksymbol (\lambda) \sim \lambda^{-4/3} {\left(\log \lambda\right)}^{1/2}$ and $w(\lambda) \sim \lambda^{-4/3} \log \lambda$~\cite{Arean:2012mq,Arean:2013tja} (see also~\cite{Ishii:2019gta}). Taking into account these considerations, 
we adopt the following Ans\"atze for $\ksymbol$ and $w$,
\begin{align}
& \ksymbol (\lambda) =  8 \,\frac{a_2 - a_1}{ 12 - x W_0 }
\left[ 1+ \frac{\scsymbol \, \kUpsymbol \lambda}{\lambda_0} + \kIRsymbol \,   \frac{e^{-\frac{\lambda_0}{\kscsymbol \lambda}}\left( 1 + \frac{\lambda_0 \kIRpsymbol}{\kscsymbol \lambda}\right){\left(\frac{\kscsymbol \lambda}{\lambda_0}\right)}^{4/3}}{\sqrt{\log\left(1 + \frac{\kscsymbol \lambda}{\lambda_0}\right)}}\right]^{-1}, \\
&  w(\lambda)= w_0 \left[1 + \frac{\scsymbol \wUpsymbol \lambda}{\lambda_0 \left( 1+ \frac{\scsymbol \lambda}{\lambda_0}\right)} + \wIRsymbol \,\frac{e^{- \frac{\lambda_0}{\wscsymbol \lambda}} \left(1+\frac{\lambda_0 \wIRpsymbol}{\wscsymbol \lambda}\right){\left( \frac{\wscsymbol \lambda}{\lambda_0}\right)}^{4/3}}{\log\left(1+\frac{\wscsymbol \lambda}{\lambda_0}\right)} \right]^{-1}.
\end{align}
In order to compute the profiles of the background fields we need to specify the values of the parameters that appear in the definition of the potentials presented above. These parameters will be fitted to the ratios between the low-spin meson masses and the $\rho^0$ meson mass as predicted by the model.

Finally, we need to specify the potentials in the CP-odd action $S_a$ \eqref{eq:axion_action}. In flat-space tachyon condensation $V^{(a)}(\lambda, \tau)$ is independent of $\lambda$ and is the same as the tachyon potential that appears in the DBI action. However in principle it may be different, so we will take $V^{(a)}$ to be $V_f$ defined above without the $V_{0f}$ term. This form guarantees that it becomes a field-independent constant at $\tau=0$ and that it vanishes exponentially at $\tau = \infty$.
The $Z(\lambda)$ function is defined by
\begin{equation}
Z(\lambda) = Z_a + c_a {\left(\frac{\lambda}{\lambda_0}\right)}^4 \, .
\label{eq:Z_func_def}
\end{equation}
The definition is constrained by Yang-Mills theory \cite{Gursoy:2007er,Gursoy:2012bt,Arean:2016hcs}. 
In this work the parameters $Z_a$ and $c_a$ will be determined by fitting the spectrum of singlet axial vector mesons.

\subsection{Evaluation of the meson spectrum} 

The quadratic fluctuations around the background fields can be mapped to the spectrum of mesons and glueballs. The normalisable fluctuations of dilaton $\Phi$, QCD axion $a$, and the (traceless part of the) metric $g_{ab}$, correspond to glueballs with $J^{PC} = 0^{++}, \, 0^{-+}$, and $2^{++}$, respectively\footnote{To be precise, it is the diffeomorphism invariant combination of the fluctuations of the dilaton and the trace of the metric which is dual to the $0^{++}$ glueballs.}. Here $J$ is the spin, $P$ refers to parity, and $C$ refers to charge conjugation. The meson sector comes from the normalisable fluctuations of the tachyon $T$ and of the gauge fields $A^{L/R}_{a}$. They correspond to mesons with $J^{PC} = 0^{++}, \, 0^{-+}, \, 1^{++},$ and $1^{--}$.  

The fluctuations can be further classified according to how they transform under the vectorial $SU(N_f)$. They can be grouped in flavour singlet and flavour non-singlet modes, i.e. mesons transforming in the adjoint of $SU(N_f)$. The fluctuations that come from $S_f$ and $S_a$ are only flavour singlet, while the ones coming from $S_g$ include singlet and non-singlet terms. The singlet terms from $S_f$ will mix with with the singlet terms coming from $S_g$ and $S_a$. 

The masses of the different glueballs and mesons can be obtained after expanding the action $S = S_g + S_f+S_a$ to quadratic order of the fluctuations of the background fields. Due to flavour and rotational covariance the fluctuations decouple in separate sectors~\cite{Arean:2013tja}, 
apart from the mixing of the flavor singlet sectors mentioned above.  
In summary, there are flavour singlet rank-two tensor fluctuations ($J^{PC} = 2^{++}$), flavour singlet and non-singlet vector mesons ($J^{PC} = 1^{--}$), flavour singlet and non-singlet axial vector mesons ($J^{PC} = 1^{++}$), flavour singlet and non-singlet scalars ($J^{PC} = 0^{++}$) and flavour singlet and non-singlet pseudoscalars ($J^{PC} = 0^{-+}$). These fluctuations generate towers of $2^{++}$ glueballs, singlet and non-singlet vector mesons, singlet and non-singlet axial vector mesons, non-singlet scalar mesons and mixtures between $0^{++}$ glueballs and $\sigma$ mesons, and non-singlet pseudoscalar mesons and mixtures between 
$0^{-+}$ glueballs and $\eta'$ mesons, respectively. 
These towers of mesons and glueballs come as solutions of a Schr\"odinger problem associated with the equation of motion of the associated fluctuation. The eigenvalues correspond to the square of the mass of the glueball or meson and their holographic wave functions are the associated eigenfunctions.
In this work we will not consider the flavour singlet states of $J^{PC} = 0^{++}, \, 0^{-+}$, as they involve mixing of the $0^{++}$ glueball with the flavour singlet $\sigma$ meson and mixing between the $0^{-+}$ glueball with the $\eta'$ meson, respectively. Therefore their analysis is considerably more challenging than that of the flavor nonsinglet states (see~\cite{Arean:2013tja,
Iatrakis:2015rga,Arean:2016hcs}) and would slow down the computer code for the spectrum significantly. Notice that these states are not central for the Regge analysis which is the main application of this work.

A detailed derivation of the equations of motion and Schr\"odinger problems associated with each fluctuation has been done in \cite{Arean:2013tja} and hence we will just summarise the main results relevant for the present  work 
(see Appendix~\ref{appendix:b} for the analysis of the spin 1 fluctuations).  
The singlet and non-singlet vector mesons have the same equation of motion
\begin{equation}
\frac{1}{V_f(\lambda, \tau) w(\lambda)^2 e^A G}\, \partial_z \Big(V_f(\lambda, \tau) w(\lambda)^2 e^A G^{-1} \partial_z \psi_V\Big) + m_V^2 \psi_V = 0 \,,
\end{equation}			
where $\psi_V=\psi_V(z)$ is their wavefunction. By performing the change of variable defined by
\begin{equation}
\frac{du}{dz} = G(z) \equiv \sqrt{1 + e^{-2A} \ksymbol (\lambda) {\left( \partial_z \tau \right)}^2} \,.
\end{equation}
and rescaling 
\begin{equation}
\psi_V (z)= \alpha(z) / \Xi_V\,, \qquad
\quad \Xi_V = w (\lambda) \sqrt{V_f (\lambda, \tau) \,e^A}\,,
\end{equation}
one can rewrite the equation of motion in the Schr\"odinger form
\begin{equation}
- \frac{d^2 \alpha}{d u^2} + V_V(u) \,\alpha = m_n^2 \,\alpha\,,
\end{equation}
 with potential
\begin{equation}
V_V (u) = \frac{1}{\Xi_V (u)} \frac{d^2 \Xi_V (u)}{d u^2}\,.
\label{eq:V_V}
\end{equation}

The singlet and non-singlet axial vector mesons have Schr\"odinger potentials differing by a term coming from the action $S_a$. The potentials of the non-singlet axial vector mesons and of the singlet axial vector mesons are, respectively,
\begin{align}
V_{NSA} (u) &= V_V(u) + 4 \,\frac{\tau^2 e^{2 A}}{w(\lambda)^2} \,\ksymbol (\lambda)\,,
 \\
V_{SA}(u) &= V_{NSA} (u) + 4 \, x \, \frac{e^{2A}Z(\lambda)V^{(a)}(\lambda, \tau)^2}{V_f(\lambda,\tau) G w(\lambda)^2}\,.
\end{align}
The non-singlet scalar mesons have the potential
\begin{equation}
V_S(u)    = \frac{1}{\Xi_S(u)} \frac{d^2 \Xi_S(u)}{d u^2} + H_S (u)\,, 
\end{equation}
with
\begin{equation}
\Xi_S(u)   = \frac{1}{G} \sqrt{V_f(\lambda, \tau) \ksymbol (\lambda) e^{3A}}\,, 
\qquad
H_S(u)     =  - \frac{e^{2A}}{\ksymbol (\lambda)} \left( \frac{(\partial_\tau V_\tau)^2}{V_\tau^2} - \frac{\partial_\tau^2 V_\tau}{V_\tau} \right),
\end{equation}
where the expression for $H_S$ differs from that of~\cite{Arean:2013tja} because our Ansatz for $V_\tau$ is different. 
Finally the equation of motion of the non-singlet pseudoscalar fluctuations is given by\footnote{Notice that the UV boundary condition for the pseudoscalar fluctuations is nontrivial and also depends on whether the quark mass is finite or not~\cite{Arean:2013tja}. A consistent way which leads to UV finiteness of the fluctuated action in all cases is to require that the factor in square brackets (rather than the wave function $\psi_P$) in~\eqref{PSflucteq} vanishes in the UV. 
For the pseudoscalars we actually solved the differential equation by using a different method than in the other sectors (i.e. by shooting) because of the complication with the boundary condition.}
\begin{align} \label{PSflucteq}
&V_f(\lambda, \tau)\, \tau^2 e^{3A} G^{-1} \ksymbol (\lambda) \, \partial_z \left[ \frac{1}{V_f(\lambda, \tau) \tau^2 \ksymbol (\lambda) e^{3A} G} \,\partial_z \psi_P \right] - \notag \\
& - 4 \tau^2 e^{2A} \frac{\ksymbol (\lambda)}{w(\lambda)^2}\, \psi_P + m^2 \psi_P = 0\,,
\end{align}
with associated Schr\"odinger potential
\begin{equation}
V_P(u) = \frac{1}{\Xi_P(u)} \frac{d^2 \Xi_P(u)}{d u^2} + H_P (u)\,,
\end{equation}
where
\begin{equation}
\Xi_P(u) = \frac{1}{\tau \sqrt{V_f(\lambda, \tau) \ksymbol (\lambda) e^{3A}}}\,, 
\qquad
H_P(u) = \frac{4 \tau^2 e^{2A} \ksymbol (\lambda)}{w(\lambda)^2}\,.
\end{equation}

The numerical determination of the spectrum proceeds by first finding the background solution (the metric and the scalar fields $\lambda$ and $\tau$) of the equations of motion defined by the action $S = S_g + S_f$. Details of the numerical procedure can be found in appendix~\ref{appendix:eom_sol}. We then solve the fluctuation equations on top of the numerical background. For the cases of singlet and non-singlet vector and axial vector fluctuations, and for  non-singlet scalar fluctuations, we compute the Schr\"odinger potential and use a pseudospectral method based on Chebyschev polynomials to compute the predicted masses of this model. The number of Chebyschev points used was 1000 and we checked the results were stable by computing the masses with a higher number of points. The reliability of the results was also studied by considering different IR and UV cutoffs on the background fields used to solve the Schr\"odinger problems. The masses of the pseudoscalars were computed using the shooting method. These methods were implemented in $C\texttt{++}$ and all results were also cross-checked against the (significantly slower) Mathematica code used in~\cite{Arean:2013tja}.

\subsection{Fitting the spectrum}\label{sec:fit}

We now proceed to fix the parameters that appear in the potentials by comparing the predictions of our model with the experimental values of the meson masses quoted by the Particle Data Group \cite{Zyla:2020zbs}. 

The overall energy units in the model is also a free parameter. Its effect on the background and  spectrum is trivial due to a scaling symmetry of the holographic model~\cite{Jarvinen:2011qe} which reflects the scale independence of the QCD Lagrangian. 
The scaling symmetry implies, in particular, that the equations of motion are unchanged under the transformation
\begin{equation}
A \rightarrow A - \log \Lambda \,, \quad z \rightarrow \Lambda z \, .
\end{equation}
By applying this transformation to the spectrum we see that all masses are scaled by the factor $\Lambda$.   
We will in effect choose $\Lambda$ such that the numerical mass of the $\rho$ meson matches the experimental result in GeV units. 
This is equivalent to fitting the parameters of the to the numerical values of ratios of masses (with respect to the $\rho$ meson mass) instead of numerical values of masses.

We will only consider  mesons made of light  up and down quarks. This sets the $x$ parameter coming from the flavour sector to be $2/3$. In table~\ref{table:light mesons} we show all the mesons listed in~\cite{Zyla:2020zbs} under \textit{light unflavoured mesons} with the values of $J^{PC}$ mentioned before. The exceptions are the flavour singlet scalars and pseudoscalars and the $a_0(980)$. Whether the latter is a quark-antiquark state or a four-quark state is still debatable, although the literature favours more the four-quark state hypothesis. For this reason we did not include it in this work. In table~\ref{table:other light mesons} we have the mesons listed  in~\cite{Zyla:2020zbs} 
under \textit{other light unflavoured mesons}, which are still not well established. 
We also included theses masses in our fit, therefore
in case some of these states are not confirmed this work should be updated.

\begin{table}
\centering
\begin{tabular}{ | c | c | c | c | }
\hline
$J^{PC}$ & I & Meson & Mass Measured (GeV) \\
\hline
$1^{--}$ & 1 & $\rho$ & 0.7755 \\
\hline
$1^{--}$ & 1 &  $\rho(1450)$ & 1.465 \\
\hline
$1^{--}$ & 1 & $\rho(1700)$ & 1.720 \\
\hline
$1^{--}$ & 0 & $\omega(782)$ & 0.78265 \\
\hline
$1^{--}$ & 0 &  $\omega(1420)$ & 1.420 \\
\hline
$1^{--}$ & 0 & $\omega(1650)$ & 1.670 \\
\hline
$1^{++}$ & 1 & $a_1(1260)$ & 1.230\\
\hline
$1^{++}$ & 0 & $f_1(1285)$ & 1.2819\\
\hline
$1^{++}$ & 0 & $f_1(1420)$ & 1.4264\\
\hline
$0^{++}$ & 1 &  $a_0(1450)$ & 1.474 \\
\hline
$0^{-+}$ & 1 &  $\pi_0$ & 0.134977 \\
\hline
$0^{-+}$ & 1 &  $\pi_0(1300)$ & 1.300 \\
\hline
$0^{-+}$ & 1 & $\pi_0(1800)$ & 1.812 \\
\hline
\end{tabular}
\caption{Light unflavoured mesons from~\cite{Zyla:2020zbs} used in the spectrum fit. The quantum number $I = 1$ means the meson is a flavour non-singlet state while $I = 0$ means the meson is a flavour singlet state.}
\label{table:light mesons} 
\end{table}

\begin{table}
\centering
\begin{tabular}{ | c | c | c | c | }
\hline
$J^{PC}$ & I & Meson & Mass Measured (GeV) \\
\hline
$1^{--}$ & 1 &  $\rho(2000)$ & 2.000 \\
\hline
$1^{--}$ & 1 & $\rho(2270)$ & 2.265 \\
\hline
$1^{--}$ & 0 &  $\omega(1960)$ & 1.960 \\
\hline
$1^{--}$ & 0 & $\omega(2205)$ & 2.205 \\
\hline
$1^{--}$ & 0 &  $\omega(2290)$ & 2.290 \\
\hline
$1^{--}$ & 0 & $\omega(2330)$ & 2.330 \\
\hline
$1^{++}$ & 1 & $a_1(1930)$ & 1.930\\
\hline
$1^{++}$ & 1 & $a_1(2095)$ & 2.095\\
\hline
$1^{++}$ & 1 & $a_1(2270)$ & 2.270\\
\hline
$1^{++}$ & 0 & $f_1(1970)$ & 1.971\\
\hline
$1^{++}$ & 0 & $f_1(2310$ & 2.310\\
\hline
$0^{++}$ & 1 &  $a_0(2020)$ & 2.025 \\
\hline
$0^{-+}$ & 1 &  $\pi_0(2070)$ & 2.070 \\
\hline
$0^{-+}$ & 1 & $\pi_0(2360)$ & 2.360 \\
\hline
\end{tabular}
\caption{Other light mesons from~\cite{Zyla:2020zbs} used in the spectrum fit. The quantum number $I = 1$ means the meson is a flavour non-singlet state while $I = 0$ means the meson is a flavour singlet state.}
\label{table:other light mesons} 
\end{table}

As our goal is to include the Regge behavior of vector and axial vector mesons, the most important criterion for the fit will be the deviation of the vector meson masses of the model from the experimental results. We have explored different fitting strategies. We tested fits where the parameters $\scsymbol$ and $\VgIRsymbol$ are fitted either independently to Yang-Mills data or together with the other parameters to ``final'' meson mass data. We also tried including lattice data for glueball masses. The result of these tests was that the optimal method, which lead to a physically sound solution for the metric and a good fit of the spin $1$ states, was to do a global simulatenous fit of all parameters, excluding the glueball masses, and also imposing specific constraints to the fit parameters. We will explain the details below.

The profile of the background fields and the mesons masses (excluding axial vector singlet states which will be discussed below) are determined by 17 parameters. 16 of these parameters ($\scsymbol$, $\kscsymbol$, $\wscsymbol$, $W_0$, $w_0$, $\kUpsymbol$, $\wUpsymbol$, $\VgIRsymbol$, $\WIRsymbol$, $\kIRsymbol$, $\wIRsymbol$, $W_1$, $\kIRpsymbol$, $\wIRpsymbol$, $a_1$, and $a_2$) are parameters of the potentials appearing in the action and $\tau_0$ is a parameter that characterises the IR asymptotics of the tachyon field. 
In our fits $a_1$ and $a_2$ are fixed by imposing the following constraints: we choose $a_2 - a_1 = 1$ by rescaling the $\tau$ field, and set 
$a_2 = 2 \ksymbol (0) $  
in order for the mass gap of the mesons to be correct at large quark mass
\cite{Jarvinen:2015ofa}. This reduces the number of free parameters to 15.

It turns out that it is useful to set extra constraints for the behavior of the tachyon which guarantee that the fit parameters remain in the domain of physically reasonable solutions. The first is related to chiral symmetry breaking. The chirally symmetric vacuum solution of the model flows to an IR fixed point~\cite{Jarvinen:2011qe}. We require that there is an instability towards forming a tachyon condensate in the IR around this fixed point, which will imply chiral symmetry breaking on the field theory side. The presence of the instability, and therefore chiral symmetry breaking, is guaranteed if the Breitenlohner-Freedman (BF) bound~\cite{Breitenlohner:1982bm} of the tachyon is violated at the fixed point. This means that $-m_\tau^2\ell_*^2>4$, where
\begin{equation}
 -m_\tau^2\ell_*^2 = \frac{24 (a_2-a_1)}{\ksymbol(\lambda_*) V_\mathrm{eff}(\lambda_*)} \, , \qquad V_\mathrm{eff}(\lambda) = V_g(\lambda) - V_f(\lambda,\tau=0) \, .
\end{equation}
Here the location of the fixed point is the maximum of the effective potential, $V_\mathrm{eff}'(\lambda_*) = 0$, and $\ell_*$ is the IR AdS radius. Actually, while violation of the BF bound guarantees tachyon condensation and chiral symmetry breaking, it turns out that, in practice, values close to the bound are enough to trigger condensation. Therefore we will in fact require $-m_\tau^2\ell_*^2\gtrsim 3.5$.

The other condition is to require that the tachyon diverges fast enough in the IR to set all potential IR boundary terms arising from the flavour action to zero. This is required, among other things, for the correct implementation of the flavour anomalies~\cite{Casero:2007ae,Arean:2016hcs}.
For our choice of potentials the asymptotics of the tachyon in the IR is
\begin{equation}
\tau \sim \tau_0 \, z^{\tau_c}\,, 
\qquad 
\tau_c =  \frac{1}{8} 
\frac{\left(12 - x W_0\right) \kIRsymbol a_2}{\VgIRsymbol (a_2 - a_1)}\,,
\end{equation} 
and the flavor action vanishes fast enough in the IR if $\tau_c>1$, which is roughly what we require below.

We then fit these parameters to ratios between the meson masses in tables~\ref{table:light mesons} and~\ref{table:other light mesons} and the $\rho$ mass, by minimising the function
\begin{equation}
J = \sum_{i} \frac{|R_{\mathrm{pred.} \,i} - R_{\mathrm{obs.} \, i}|}{R_{\mathrm{obs.} \, i}} + \lambdapar \,e^{- \left(4 \tau_c/3 - 1\right)} + \lambdapar  \,e^{-\left(  -m_\tau^2\ell_*^2 - 3.5 \right)},
\end{equation} 
excluding the singlet axial vector mesons. This gives a total of 22 data points. The sum term is the absolute relative difference between the predictions of our model and the ones obtained by using experimental data, while the other two terms are the constraints we want our background to satisfy. 
We repeated the fit with different values of the $\lambdapar$ parameter in order to balance the ability of the model to reproduce the observed ratios and still be consistent and stable. We have found that $\lambdapar = 0.1$ is a good choice and the results that we present below were obtained with such value. Having fixed the background, we fit the parameters $Z_a$ and $c_a$ of equation~\eqref{eq:Z_func_def} against the four mass ratios between the singlet axial vector mesons and the $\rho$ meson. With this procedure we have obtained the parameter values presented in table~\ref{table:best_fit_background_pars} and the corresponding mass ratios of table~\ref{table:best spectrum fit}. 
This fit has $- m_\tau^2 \ell_*^2 \approx 6.200$ and $\tau_c \approx 1.956$ which ensure presence of chiral symmetry breaking and IR decoupling of the tachyon.  
\begin{table}
\centering
\begin{tabular}{|c|c|c|c|c|c|}
\hline
Parameter & value & Parameter & value & Parameter & value \\
\hline
$\scsymbol$ & 2.8328 & $\kscsymbol$ & 3.1670 & $\wscsymbol$ & 1.6926 \\
\hline
$W_0$ & 2.4289 & $w_0$ & 0.9400 & $\kUpsymbol$ & 1.3254 \\ 
\hline
$\wUpsymbol$ & -0.2898 & $\VgIRsymbol$ & 1.8042 & $\WIRsymbol$ & 1.1345 \\ 
\hline
$\kIRsymbol$ & 1.7647 & $\wIRsymbol$ & 2.9291 & $W_1$ & 0.2342 \\ 
\hline
$\kIRpsymbol$ & -0.3076 & $\wIRpsymbol$ & 3.0358 & $a_1$ & 0.5413 \\
\hline
 $a_2$ & 1.5413 & $\tau_0$ & 0.9232 & $Z_a$ & -0.0377 \\
\hline
$c_a$ & 23.307 & & & & \\
\hline
\end{tabular}
\caption{Best fit parameters of the background potentials to the mass ratios between the mesons listed on tables~\ref{table:light mesons} and~\ref{table:other light mesons} and the $\rho$ meson. }
\label{table:best_fit_background_pars} 
\end{table}
\begin{table}
\centering
\begin{tabular}{ | c | c | c | c | }
\hline
Ratio & $\mathrm{R_{pred.}}$ & $\mathrm{R_{obs.}}$ & $|\mathrm{R_{pred.}} - \mathrm{R_{obs.}}| / \mathrm{R_{obs.}}$ \\
\hline
$m_{\rho(1450)}/m_\rho$ & $1.662$ &	$1.890$ & $0.121$ \\
\hline
$m_{\rho(1700)} / m_\rho$ & $2.141$ & $2.219$ & $0.035$ \\
\hline
$m_{\rho(2000)} / m_\rho$ & $2.559$ & $2.580$ & $0.008$ \\
\hline
$m_{\rho(2270)} / m_\rho $ & $2.940$ & $2.922$ & $0.006$ \\
\hline
$m_{\omega(782)} / m_{\rho}$ & $1$ & $1.010$ & $0.010$ \\
\hline
$m_{\omega(1420)} / m_\rho$ & $1.662$ & $1.832$ & $0.093$ \\
\hline
$m_{\omega(1650)} / m_{\rho}$ & $2.141$ & $2.154$ & $0.006$ \\
\hline
$m_{\omega(1960)} / m_\rho$ & $2.559$ &	$2.528$ & $0.012$ \\
\hline
$m_{\omega(2205)} / m_{\rho}$ & 2.940 &2.844 &0.034 \\
\hline
$m_{\omega(2290)} / m_{\rho}$ & $3.289$ & $2.954$ & $0.127$\\
\hline
$m_{\omega(2330)} / m_{\rho}$ & $3.610$ & $3.005$ &	$0.201$ \\
\hline
$m_{a_1(1260)} / m_\rho$ & $1.591$ &	 $1.587$ & $0.003$ \\
\hline
$m_{a_1(1930)} / m_\rho $ & $2.095$ & $2.489$ &	$0.158$ \\
\hline
$m_{a_1(2095)} / m_\rho$ & $2.523$ &	 $2.702$ & $0.066$ \\
\hline
$m_{a_1(2270)} / m_\rho$ & $2.916$ & $2.928$ & $0.004$ \\
\hline
$m_{f_1(1285)} / m_\rho$ & $1.653$ &	$1.654$ & $0.001$ \\
\hline
$m_{f_1(1420)} / m_{\rho}$ & $2.128$ & $1.840$ &  $0.157$ \\
\hline
$m_{f_1(1970)} / m_{\rho}$ & $2.543$ &	$2.542$ & $0.0004$ \\
\hline
$m_{f_1(2310)} / m_{\rho}$ & $2.930$ & $2.980$ & $0.017$ \\
\hline
$m_\pi / m_\rho$ & $0.1740$ & $0.1741$ & $0.0006$ \\
\hline
$m_{\pi(1300)} / m_\rho$ & $1.731$ & $1.677$ & $0.032$ \\
\hline
$m_{\pi(1800} / m_\rho$ & $2.337$ & $2.337$ & $5 \times 10^{-5}$ \\
\hline
$m_{\pi(2070)} / m_\rho$ & $2.785$ & $2.670$ & $0.043$ \\
\hline
$m_{\pi(2360)} / m_\rho$ & $3.173$ & $3.044$ & $0.042$ \\
\hline
$m_{a_0(1450)} / m_\rho$ & $0.685$ & $1.901$ & $0.640$ \\
\hline
$m_{a_0(2020)} / m_\rho$ & $1.492$ & $2.612$ & $0.429$ \\
\hline
\end{tabular}
\caption{Mass ratios obtained with the parameter values of table~\ref{table:best_fit_background_pars}.}
\label{table:best spectrum fit} 
\end{table}

Several remarks are in order. Firstly, the fit is stiff: while the number of parameters is large, the dependence of the results on their values is relatively mild. This is because the fit parameters appear through only a few functions of $\lambda$, the asymptotics at large and small coupling of which have already been determined  by qualitative arguments and comparison to perturbation theory. Therefore the fit parameters essentially only affect the functions in the middle, at $\lambda \sim 1$. Also there is limited parameter space where the functions are simple, monotonic functions, and one can check from the fit result that it indeed lies within this regime of the parameter space. 

Given the stiffness of the fit, the results for the spin-one mesons are really good. There are a few isolated states for which the deviation is $\gtrsim 10\%$, but in general the deviations are in the ballpark of $1\%$ or even less than that. The masses of the pseudoscalar mesons are also reproduced at a very good precision. There are, however, significant deviations in the scalar sector.  While the scalar sector is challenging to explain in any model among other things due to the presence of significant four-quark contribution~\cite{Zyla:2020zbs}, the predicted nonsinglet scalar masses are still clearly too low, unlike in the probe limit study of~\cite{Iatrakis:2010zf,Iatrakis:2010jb} which used similar flavor action as the current article with the fixed background of~\cite{Kuperstein:2004yf}. 
While exploring different fit procedures, we noticed that there are parameter values for which the scalar masses are reproduced to a much better precision, but such parameter values are not favored by the overall fit which stresses the masses of the spin 1 mesons. 
Understanding this shortcoming requires further study. Notice that the scalar states are not needed for the analysis of the Regge trajectories which is the topic of discussion in the remainder of this paper. 

The fitted value of $Z_a = Z(0)$ in Table~\ref{table:best_fit_background_pars} is negative. Due to the positive value of $c_a$, however, the function $Z(\lambda)$ is mostly positive so that $Z(\lambda)$ has a node at small $\lambda$. This behavior is unexpected and does not agree with phenomenology, i.e. the physics of the $\theta$-angle in QCD and in particular the value of the topological susceptiblity~\cite{Gursoy:2007er,Arean:2016hcs}. Apparently this issue arises because the singlet axial meson masses are relatively unsensitive to the shape of the function $Z(\lambda)$, and would be cured if additional observables (e.g. the topological susceptiblity) would be included in the fit. The precise functional form of $Z(\lambda)$ is again irrelevant for the Regge analysis of the following sections. 

It is also interesting to compare the fit results to those obtained in the same holographic  
model~\cite{Jokela:2018ers} by fitting the potentials independently to lattice data for QCD thermodynamics. Namely, most of the values are very close to those obtained in that study, deviations are typically in the ballbark of $10\%$. In particular, the scale parameters $\scsymbol$ and $\kscsymbol$ are close to the value $1$ used in this reference, whereas $\wscsymbol$ is a bit higher than what was obtained through the fit to thermodynamics ($\wscsymbol$ equals to $3w_s$ of~\cite{Jokela:2018ers}) but still smaller than the other scale parameters, in agreement with the earlier fit.   
We also note that there is rough agreement with~\cite{Gursoy:2016ofp,Gursoy:2020kjd} where the model was compared to lattice results at finite magnetic field and temperature by using a sligthly different Ansatz for the potentials of the model. The parameter $c$ of these references, which controls the scale in the $\lambda$ dependence of the $w(\lambda)$ function, maps roughly to the ratios $\wscsymbol/\scsymbol$ or $\wscsymbol/\kscsymbol$ in this article. For the ratios we obtain numbers close to $0.5$ whereas $c \approx 0.25$ was preferred by the thermodynamics at finite magnetic field. That is, the numerical values are different, but clearly smaller than one in both cases. Moreover the value of $W_0$ (which was a free parameter in~\cite{Jokela:2018ers}) is determined by the spectrum fit to be near 2.5. This results is therefore an important constraint with respect to the earlier fit. The most significant difference between the fits is the value of $\bar w_0$, which here is  smaller by a factor of about 5 to 10 with respect to the various fits of~\cite{Jokela:2018ers}. This is apparently connected to the change in the value of $\wscsymbol$. 

\section{$\gamma\gamma$, $\gamma p$ and $pp$ total cross-sections in holographic QCD}
\label{seq:hol_xsections}

In this section we present the necessary ingredients to compute the total cross-sections of $\gamma\gamma$, $\gamma p$ and $pp$ scattering in holographic models of QCD in the Veneziano limit. First we will discuss the kinematics of each process. Then we will present generic holographic expressions of the forward scattering amplitude, in the Regge limit, via the exchange of higher spin $J$ fields. We conclude by deriving the holographic expression of the total cross-sections by taking the imaginary part of the amplitudes and using the optical theorem.

\subsection{Kinematics}

For all processes we will use light-cone coordinates $\left(+,-,\perp \right)$, with flat space metric $ds^2 = - dx^+ dx^- + d x^2_\perp$, where $x_\perp \in \mathbb{R}^2$.
For the $\gamma^{*} p \rightarrow \gamma^{*} p $ process the
  incoming and outgoing off-shell photons are the following
\begin{equation}
k_1=\left(\!\sqrt{s},-\frac{Q_1^2}{\sqrt{s}} ,0\right),
\ \ \ \ \ 
-k_3=\left(\sqrt{s} , \frac{q_\perp^2 - Q_3^2}{\sqrt{s}}, q_\perp \right),
\label{eq:gp_off_shell_photons_kinematics}
\end{equation}
while the incoming and outgoing protons with mass $M$ have momenta
\begin{equation}
k_2=\left(\frac{M^2}{\sqrt{s}} , \sqrt{s}, 0\right),
\ \ \ \ \ 
-k_4=\left(\frac{q_\perp^2 + M^2}{\sqrt{s}}, \sqrt{s},  -q_\perp \right).
\label{eq:gp_off_shell_protons_kinematics}
\end{equation}
The momentum transfer $q_\perp$ is a $\mathbb{R}^2$ vector and is related to the Mandelstam variable $t$ through $t = - q_\perp^2$.
We work in the Regge limit of large Mandelstam variable $s$.
For the forward scattering amplitude the 
momentum transfer $q_\perp = 0$ and the photon virtualities satisfy 
$Q_3 = Q_1 = Q$, since  the outgoing off-shell photon has $k_3=-k_1$ and the outgoing
proton $k_4=-k_2$. The  incoming and outgoing photon polarizations are the same.
The possible polarization vectors are
\begin{equation}
  \label{eq:polarization vectors} 
n(\lambda) = \begin{cases}
    \big(0,0,\epsilon_\lambda\big) \,, & \ \ \ \lambda=1,2 \\
   \big( \sqrt{s}/Q, Q/\sqrt{s},0  \big)\, , & \ \ \ \lambda=3
    \end{cases}
    \,,
\end{equation}
where $\epsilon_\lambda$ is just the usual transverse polarization vector.

For the $\gamma^{*} \gamma^{*} \rightarrow \gamma^{*}  \gamma^{*} $ process the 
incoming photons   have the four momenta
\begin{equation} 
k_1 = \left( \sqrt{s}, - \frac{Q_1^2}{\sqrt{s}}, 0 \right) \,,  \qquad  
k_2 = \left( - \frac{Q_2^2}{\sqrt{s}}, \sqrt{s},  0 \right) \,,
\end{equation}
while the outgoing photons have
\begin{equation}
k_3 = - \left( \sqrt{s},  \frac{q_\perp^2 - Q_3^2}{\sqrt{s}}, q_\perp \right) \, \qquad  k_4 = - \left(  \frac{q_\perp^2-Q_4^2}{\sqrt{s}}, \sqrt{s},  -q_\perp \right),
\end{equation}
where $Q^2_i = k_i^2 >0$  $(i=1, \dots, 4)$  are the corresponding virtualities. As in the case of $\gamma^* p$ scattering, for the forward scattering amplitude the 
momentum transfer is null.
The possible polarization vectors are, respectively,
\begin{align}
    &n_{1,3} =
    \begin{cases}
      \left(0,0,1,0\right), & \lambda=1 \\
      \left(0,0,0,1\right), & \lambda=2 \\
      \frac{1}{Q} \left( \sqrt{s}, \frac{Q^2}{\sqrt{s}}, 0, 0 \right), & \lambda = 3
    \end{cases}\,,\\
    &n_{2,4} =
    \begin{cases}
      \left(0,0,1,0\right), & \lambda=1 \\
      \left(0,0,0,1\right), & \lambda=2 \\
      \frac{1}{Q} \left( \frac{Q^2}{\sqrt{s}}, \sqrt{s}, 0, 0 \right), & \lambda = 3
    \end{cases} \,,
    \label{eq:inPolarization}
\end{align}
since in the forward amplitude the incoming and outgoing off-shell photons have the same polarizations.
Notice that the transverse photons~$\left(\lambda=1,2\right)$ are normalized such that~$n^2 = 1$, while for the  longitudinal photons~$\left(\lambda=3\right)$  $n^2 = -1$.

Finally, the large $s$ kinematics of $pp$ scattering is given by
\begin{align}
\label{eq:pp_kinematics}
&k_1=\left(\!\sqrt{s},\frac{M^2}{\sqrt{s}} ,0\right),\  \ k_3=-\left(\!\sqrt{s},\frac{ q_\perp^2 +M^2}{\sqrt{s}} , q_\perp \right)\!,\\
&k_2=\left(\frac{M^2}{\sqrt{s}},\sqrt{s} ,0\right),\  \ k_4=-\left(\frac{M^2+ q_\perp^2}{\sqrt{s}},\sqrt{s} ,-q_\perp \right), \notag
\end{align}
where $k_1$ and $k_2$ are the incoming proton momenta and $k_3$ and $k_4$ are the outgoing proton momenta. As in the other processes we will only compute the forward scattering amplitude for which $q_\perp = 0$.

\subsection{Holographic scattering amplitudes}
\begin{figure}
\begin{center}
\subfloat[]{ \includegraphics[height = 4cm]{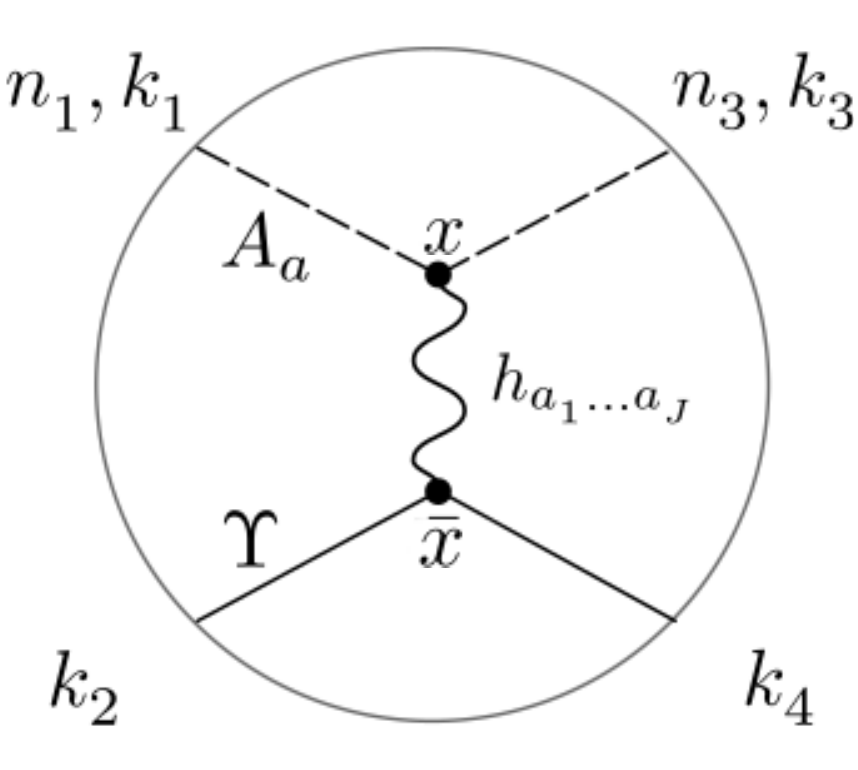} } 
\subfloat[]{ \includegraphics[height = 4cm]{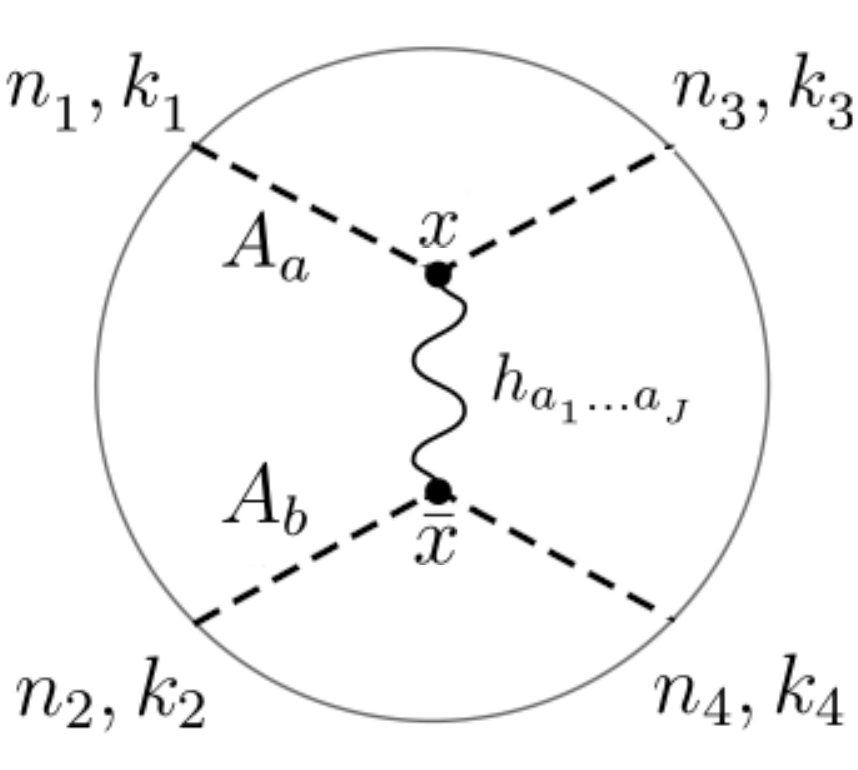} }
\subfloat[]{ \includegraphics[height = 4cm]{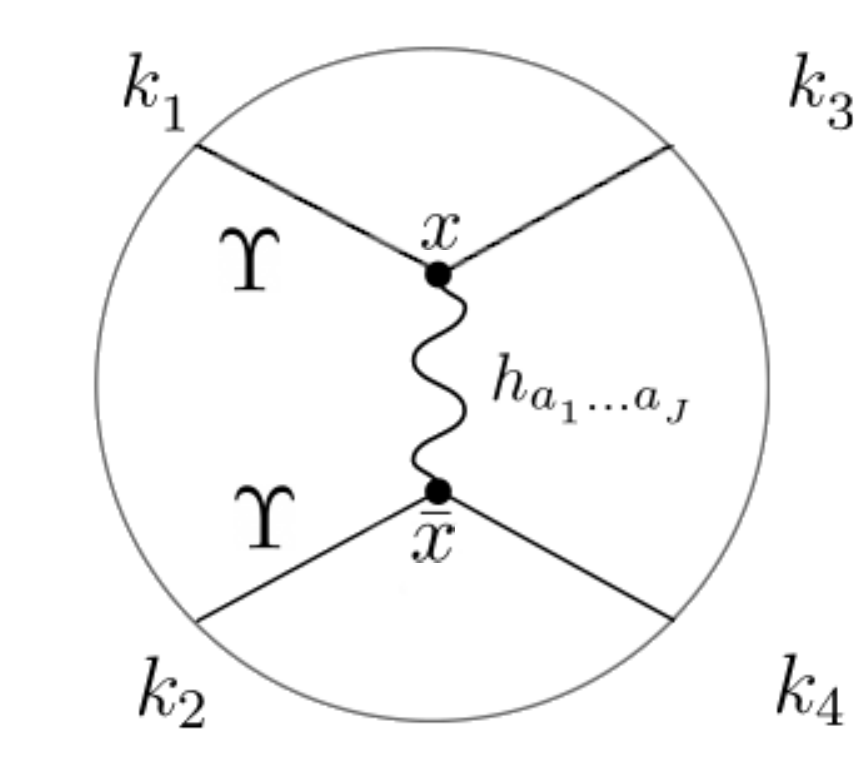} }
\end{center}
\caption{Tree level Witten diagram representing spin $J$    exchange in  (a) $\gamma^* p\to\gamma^* p$, (b) $\gamma \gamma \to \gamma \gamma$ and (c) $p p\to p p$ scattering. 
The $n_1$ and $n_2$ labels denote the incoming photon polarizations while $n_3$ and $n_4$ label the outgoing photon polarizations. For forward scattering $n_1=n_3$ and $n_2 = n_4$. $A_a$ represents the non-normalizable mode of a $U(1)$ gauge field dual to the source of the conserved current $\bar{\psi} \gamma^\mu \psi$ and $\Upsilon$ is a normalizable mode of a bulk scalar field that represents an unpolarised proton. $x$ and $\bar{x}$ represent the bulk points where the external scattering states couple with the spin $J$ fields.}
\label{fig:Witten_diagrams}
\end{figure}
Before we start the computation of the forward scattering amplitudes we need to define the external states as well as the interaction between them and the spin $J$ fields that are exchanged in the Witten diagrams of figure~\ref{fig:Witten_diagrams}.

An external photon is a source of the conserved current $\bar{\psi} \gamma^\mu \psi$, where the quark field $\psi$ comes from the open string sector. According to the gauge/gravity duality this field is dual to the nonnormalizable mode of a vector field in the bulk. In the context of this model the natural candidate is the linear combination of the $A^L$ and $A^R$ gauge fields
\begin{align}
V_a = \frac{A^L_a + A^R_a}{2} \,.
\end{align}
In the string frame the action of this field is
\begin{equation}
S = - \frac{1}{4} M^3 N_c N_f \int d^5 x \sqrt{-g_s} e^{-\frac{10}{3} \Phi} V_f \,w_s^2\, G F_{ab}\tilde{g}^{ac} \tilde{g}^{bd}F_{cd} \, ,
\label{eq:VM_new_action_string_frame}
\end{equation}
where $g_s$ is the determinant of the metric in the string frame, $\tilde{g}^{ab}$ is the inverse of $\tilde{g}_{ab} = g_{\mathrm{s} ab} +  \ksymbol _s(\lambda) \partial_a \tau \partial_b \tau $, $\ksymbol _s(\lambda) = \lambda^{4 / 3} \ksymbol(\lambda) $ and $w_s (\lambda) = \lambda^{4 / 3} w(\lambda)$. Working in the gauge $V_z = 0$ and $\partial_\mu V^{\mu} = 0$, the vector field components describing a boundary plane wave solution with polarization $n_\mu$ take the form
\begin{equation}
V_\mu \left(x, z\right) = n_\mu f_Q \left(z\right) e^{i k \cdot x} \, , \qquad k^2 = Q^2 \, ,
\end{equation}
where $f_Q$ satisfies the differential equation
\begin{equation} \label{eq:fQeq}
\frac{1}{V_f(\lambda, \tau) w(\lambda)^2 e^A G} \, \partial_z \Big(V_f(\lambda, \tau) w(\lambda)^2 e^A G^{-1} \partial_z f_Q\Big) - Q^2 f_Q = 0 \, ,
\end{equation}
subject to the boundary conditions $f_Q\left(0\right) = 1$ and $\partial_z f_Q \left( z \to \infty \right) = 0$.
For the computation of the Witten diagrams that have photons as external states, in particular to compute the bulk interaction vertex,
it is convenient to know the field strength of a given mode
\begin{equation}
F_{\mu \nu} = 2 i k_{[ \mu} n_{\nu ]} f_Q (z) e^{i k \cdot x} \, , \quad F_{z \mu} = n_\mu \dot{f_Q} (z)e^{i k \cdot x} \, ,
\label{eq:strength_tensor}
\end{equation}
where we use the notation $\dot{f_Q} = \partial_z f_Q$.

For the proton external state we consider that it is dual to the normalizable mode of a bulk scalar field $\Upsilon\left(x, z\right) = e^{i P \cdot x} \upsilon\left(z\right)$ that represents an unpolarised proton. We will see that the contribution of the proton wavefunction to the scattering amplitudes is inside an integral that will be absorbed in the coupling constants and hence the precise details will not be important.

The last ingredient of our model are the higher spin fields $h_{a_1 \cdots a_J}$ that will mediate the interaction between the external states in the considered scattering states. In this work we will consider bulk spin $J$ fields that are dual to the spin $J$ twist two operators made of the gluon field,  as well as bulk spin $J$ fields dual to the spin $J$ twist two operators made of the quark bilinears. This extends the previous works~\cite{Ballon-Bayona:2015wra, Ballon-Bayona:2017vlm, Amorim:2018yod}, where only  bulk fields dual to the gluon operators were considered. As discussed in appendix~\ref{appendix:b}, we will consider a coupling between the $U(1)$ gauge field and these spin $J$ fields given by
\begin{align}
k_J \int d^5x \sqrt{-g_s} G e^{-\frac{10}{3} \Phi} V_f  ( \lambda, \tau)\, {w_s (\lambda )}^2  \tilde{g}^{ab} F^V_{a c} \nabla_{a_1} \dots  \nabla_{a_{J-2}} F^V_{b d} \, h^{c d a_1 \dots a_{J-2}} \, ,
\label{eq:gauge_field_spin_J_coupling}
\end{align}
while for the scalar field $\Upsilon$ dual to the proton state the coupling is
\begin{align}
\bar{k}_J \int d^5 x \sqrt{-g_s} \,e^{-\Phi} \left( \Upsilon  \nabla_{b_1} \dots  \nabla_{b_J} \Upsilon \right) h^{b_1 \dots b_J} \, .
\label{eq:proton_spin_J_coupling}
\end{align}
In equations (\ref{eq:gauge_field_spin_J_coupling}) and (\ref{eq:proton_spin_J_coupling}) $ \nabla_a$ is the covariant derivative, while $k_J$ and $\bar{k}_J$ are the couplings constants between the the $U(1)$ gauge field and the bulk scalar with the spin $J$ field, respectively.

The higher spin $J$ field $h_{a_1 \cdots a_J}$ is totally symmetric, traceless and satisfies the transversality property $ \nabla^{a_1} h_{a_1 \cdots a_J} = 0$.
This implies that, in the Regge limit, it is not important in which external fields the covariant derivatives in~\eqref{eq:gauge_field_spin_J_coupling} and~\eqref{eq:proton_spin_J_coupling} act. 
Below we assume that the spin $J$ field has a propagator, without specifying its form. In the next section  we focus on the dynamics of this field in detail for the case of pomeron and meson trajectories.

Now we will show how to compute the forward scattering amplitude for the case of $\gamma p$ scattering. The calculations for $\gamma \gamma$ and $pp$ follow the same pattern and bring no additional
dificulty. For those cases we will simply present the results. In the Regge limit, the amplitude describing the spin $J$ exchange between the incoming gauge field $V_a^{(1)} \sim  e^{i k_1 \cdot x}$ and scalar field $\Upsilon^{(2)} \sim  e^{i k_2 \cdot x}$, and outgoing gauge field $V_a^{(3)} \sim  e^{i k_3 \cdot x}$ and scalar field $\Upsilon^{(4)} \sim  e^{i k_4 \cdot x}$, can be written as
\begin{align}
&\mathcal{A}_J = k_J \bar{k}_J \int d^5 x \int d^5 \bar{x} \sqrt{-g_s} \sqrt{-\bar{g}_s} \,G \,e^{-\frac{10}{3} \Phi} V_f \,w_s^2 \,e^{-\bar{\Phi}}  \times \notag \\ 
& \times \tilde{g}^{ab} F^{(1)}_{a - } (x) \partial_{-}^{J-2}F^{(3)}_{b -} (x) \,\Pi^{- \cdots -, + \cdots +}(x,\bar{x})  \Upsilon^{(2)}(\bar{x}) \partial_{+}^J \Upsilon^{(4)}(\bar{x}) \, ,
\end{align}
where bars denote quantities evaluated at $\bar{x}$ in the Witten diagrams. 
The tensor $\Pi^{a_1 \cdots a_J, b_1 \cdots b_J}(x,\bar{x}) $ is the propagator of the spin $J$ field. Using the kinematics of equations~\eqref{eq:gp_off_shell_photons_kinematics},~\eqref{eq:gp_off_shell_protons_kinematics} and~\eqref{eq:polarization vectors}, the expressions~\eqref{eq:strength_tensor} and summing over the photon polarisations, the amplitude takes the form
\begin{align}
  \mathcal{A}_J = k_J \bar{k}_J s^J & \int d^5x d^5 \bar{x} \sqrt{-g_s} \sqrt{-\bar{g}_s}  \,G \,e^{-\frac{10}{3}\Phi} V_f \,w_s^2 \,e^{-\bar{\Phi}} e^{-2 J \left(A+\bar{A}\right)}  e^{-2A}  \times \notag \\
  &\times \left(  {f_Q}^2 + \frac{\dot{f}_Q^{\ 2}}{Q^2 G^2}  \right) {\bar{\upsilon}}^2 e^{-i q_\perp \cdot \left( x_\perp - \bar{x}_\perp \right)} \Pi _{+\dots+,-\dots-}\left(x, \bar{x} \right) \, .
\end{align}
To make progress we make the change of variable $ x - \bar{x} = (w^+, w^-, l_\perp) \equiv w$ and define the transverse propagator in transverse space $G_J (z, \bar{z}, t)$ through
\begin{align}
 \int d^2 l_\perp e^{- i q_\perp l_\perp} \int \frac{dw^+ dw^-}{2} \Pi_{+ \dots +, - \dots -} \left(z, \bar{z}, w^+, w^-, l_\perp \right) = - \frac{i}{2^J} {\left( e^{A + \bar{A}} \right)}^{J-1}G_J (z, \bar{z}, t) \, ,
\label{eq:gJ_t_def}
\end{align}
that is valid both for spin $J$ fields of the graviton Regge and meson trajectories. Defining $V = {\left(2 \pi\right)}^4 \delta^4\left( \sum k_i \right)$ we obtain
\begin{align}
  \mathcal{A}_J = - i V \,\frac{k_J \bar{k}_J}{2^J} s^J & \int dz d\bar{z} \,e^{2 A}e^{4 \bar{A}} \,G\, e^{-\frac{10}{3}\Phi} \,V_f \, w_s^2 \, e^{-\bar{\Phi}} e^{-J \left(A+\bar{A}\right)} \times \notag \\
 & \times \left(  {f_Q}^2 + \frac{ \dot{f}_Q^{\ 2}}{Q^2 G^2}  \right)  {\bar{\upsilon}}^2 G_J (z, \bar{z}, t) \, .
\end{align}
In the next section we will propose phenomenological equations of motion for the higher spin fields $h_{a_1 \cdots a_J}$ of both trajectories. In particular, it will be shown that the function $G_J (z, \bar{z}, t)$ for the pomeron and meson trajectories admits a spectral decomposition associated to a Schr\"odinger potential that describes spin $J$ glueballs or spin $J$ mesons. This function can be written in terms of the eigenfunctions $\psi_n (J, z)$ and eigenvalues $t_n (J)$ of this Schr\"odinger potential in the following way
\begin{align} 
G_J (z , \bar{z}, t ) = e^{B + \bar{B}}  \sum_n \frac{\psi_n(J, z) \,\psi_n^* (J, \bar{z})}{t_n(J) - t} \, .
\end{align}
The function $B(z)$ depends on the holographic QCD model as well on the trajectory that the higher spin field belongs to. We will determine them in the next section.

Finally, 
in order to get the total amplitude we need to sum over the spin $J$ fields with $J \geq J_{min}$, where $J_{min}$ is the minimal spin in the corresponding Regge trajectory. Then we can apply a Sommerfeld-Watson transform
\begin{align}
\frac{1}{2} \sum_{J \geq J_{min}} \left(s^J + {\left(-s\right)}^J\right) \frac{\mathcal{A}_J}{s^J} = - \frac{\pi}{2} \int \frac{d J}{2 \pi i} \frac{s^J + \left(-s\right)^J}{\sin \pi J} \frac{\mathcal{A}_J}{s^J}\,,
\end{align}
which requires analytic continuation of the amplitude for the spin $J$ exchange to the complex $J$-plane. The contour on the complex plane consists of circles around simple poles at integer values of $J$. Then, we assume that the $J$-plane integral can be deformed from the poles at 
physical values of $J$
to the poles $J = j_n (t)$ defined by $t_n (J) = t$. The scattering domain of negative t contains these poles along the real axis for $J < J_{min}$. The scattering amplitude for $t = 0$ is then
\begin{equation}
\mathcal{A} (s, 0 ) = \sum_n g^{\gamma p}_n s^{j_n\left(0\right)} \int dz \, e^{A(2 - {j_n\left(0\right)})} \,G\, e^{-\frac{10}{3} \Phi}  \,V_f \, w_s^2   \left(  {f_Q}^2 + \frac{ \dot{f}_Q^{\ 2}}{Q^2 G^2}  \right) e^{B} \psi_n (j_n, z) \,, 
\label{eq:gp_forward_scattering_amp}
\end{equation}
where
\begin{equation} 
g_n^{\gamma p} = \frac{\pi}{2} \frac{k_{j_n\left(0\right)} \bar{k}_{j_n\left(0\right)}}{2^{j_n\left(0\right)}}  \left(i + \cot \frac{\pi j_n\left(0\right)}{2} \right) \frac{d j_n}{dt} 
\int d\bar{z} \, e^{\bar{A}\left( 4 - j_n \right)} e^{-\bar{\Phi}}  {\bar{\upsilon}}^2 e^{\bar{B}}  {\psi (j_n, \bar{z})}^{*} \, .
 \label{eq:gp_couplings_def}
\end{equation}
By analysing (the regular solution to) the equation~\eqref{eq:fQeq} we see that 
\begin{equation}
\lim_{Q\rightarrow 0} f_Q = 1 \,, \qquad \lim_{Q\rightarrow 0} \frac{\dot{f}_Q}{Q} = 0\,.
\end{equation}
It therefore
follows from the optical theorem that
\begin{equation}
\sigma(\gamma p \rightarrow X) = \sum_n {\mathrm{Im}}  \big( g^{\gamma p}_n \big)   s^{j_n - 1} \int du \, e^{-\left(j_n - 2 \right)A } e^{-\frac{10}{3} \Phi} \,V_f \, w_s^2  \, e^B \psi_n(u) \, ,
\label{eq:sigma_gp_hol_gens}
\end{equation}
where we have made the change of variable $d u = G d z$.

As mentioned before the procedure to compute the holographic total cross-sections for the other processes  is similar to the one just presented. These calculations are done in appendices~\ref{sec:pp_scattering} and~\ref{sec:gg_scattering}.
Hence, we finish this section by presenting the final results. For  $ \gamma \gamma \to X $ we have 
we have 
\begin{equation}
\sigma(\gamma \gamma \rightarrow X) =  \sum_n {\mathrm{Im}} \big( g^{\gamma \gamma}_n \big) s^{jn-1} \int du \,e^{-\left(j_n - 2 \right)A } e^{-\frac{10}{3} \Phi} \,V_f \, w_s^2  \, e^B \psi_n (u)  \, , \label{eq:sigma_gg_hol_gens} 
\end{equation}
where
\begin{equation}
g^{\gamma \gamma}_n = \frac{\pi}{2} \frac{k^2_{j_n\left(0\right)}}{2^{j_n\left(0\right)}}  \left(i + \cot \frac{\pi j_n\left(0\right)}{2} \right) \frac{d j_n}{dt}  \int d\bar{u}  \, e^{-\left(j_n - 2 \right)\bar{A} } e^{-\frac{10}{3} \bar{\Phi}} \,\bar{V_f}  \, \bar{w}_s^2  \, e^{\bar{B}} {\psi_n(u)}^{*} \, . \label{eq:gg_couplings_def} 
\end{equation}
For  the $p p \to X$ process the total cross-section is  
\begin{equation}
\sigma( p p \to X ) = \sum_n {\mathrm{Im}} \big( g^{p p}_n \big) s^{jn-1}  \, , \label{eq:sigma_pp_hol_gens} 
\end{equation}
where
\begin{equation}
g^{p p}_n = \frac{\pi}{2} \frac{\bar{k}^2_{j_n\left(0\right)}}{2^{j_n\left(0\right)}}  \left(i + \cot \frac{\pi j_n\left(0\right)}{2} \right) \frac{d j_n}{dt} {\left| \int dz\,  e^{A\left( 4 - j_n \right)} e^{-\Phi}  {\bar{\upsilon}}^2 e^{B}  {\psi(j_n, z)} \right|}^2  \, \label{eq:pp_couplings_def} .
\end{equation}

\section{Spin $J$ Dynamics}
\label{seq:spin_j_dynamics}

The Witten diagrams of figure~\ref{fig:Witten_diagrams} allow to compute  four-point functions  dominated by exchange of   twist 2 operators in the large $s$ limit. We consider spin $J$ operators $\mathcal{O}_J$ of the form
\begin{equation}
\mathcal{O}_J \sim \mathrm{tr} \left[ F_{\beta \alpha_1} D_{\alpha_2} \cdots D_{\alpha_{J - 1}} F^{\beta}_{\, \, \alpha_J} \right] \, , 
\qquad
\mathcal{O}_J \sim \, \mathrm{tr} \left[ \bar{\psi} \gamma_{\alpha_1} D_{\alpha_2} \cdots D_{\alpha_J} \psi \right] . 
\end{equation}
The first set of operators are gluonic   while the second set are quark  operators.
These twist 2 operators are dual to bulk spin $J$ fields whose dynamics will be specified below. We shall then follow an effective field theory approach, by proposing a general form of the equations of motion with phenomenological parameters that can be fixed by data. We will propose two different equations of motion, one that describes the gluon sector, which includes the energy-momentum tensor $T_{\alpha\beta}$,  and the other the quark sector, which includes the quark bilinear current $J_\alpha$. These equations will satisfy two requirements: $i)$ compatibility with the graviton's equation of motion for the case $J = 2$ for gluonic operators, and compability for $J=1$ with the equation of motion of the $U(1)$ current dual to the operator $J_\alpha$; $ii)$ reduction to the conformal limit case (pure AdS space and constant dilation and tachyon).

In pure AdS the equation of motion of a spin $J$ field is
\begin{equation}
\left( \nabla^2 - M^2 \right) h_{a_1 \cdots a_J} = 0 \, , \qquad {\left(LM\right)}^2 = \Delta \left(\Delta - 4 \right) - J \, , 
\label{eq:spin_J_eom_ads}
\end{equation}
where $L$ is the AdS length scale and $\Delta$ is the dimension of the dual operator $\mathcal{O}_J$. These fields are symmetric, traceless and transverse ($\mathrm{TT}$). 
The independent components are the ones along the boundary direction (i.e. $h_{\alpha_1 \cdots \alpha_J}$) due to the transversality condition. 
Decomposing these into irreducible representations of the Lorentz group $SO(1,3)$, the $\mathrm{TT}$ components $h^{\mathrm{TT}}_{\alpha_1 \cdots \alpha_J}$ decouple from the others and they describe the operator $\mathcal{O}_J$ in the dual theory. The UV asymptotics of these fields are
\begin{align}
h_{\alpha_1 \cdots \alpha_J} \sim z^{4 - \Delta - J} \mathcal{J} + \cdots + z^{\Delta - J} \langle \mathcal{O}_J \rangle + \cdots\,,
\label{eq:asymptotics}
\end{align}
with the free theory value $\Delta=J+2$, where 
$\mathcal{J}$ is the source of $\mathcal{O}_J$ and $\langle \mathcal{O}_J \rangle$ its vacuum expectation value.

We will now motivate two equations of motion for spin $J$, one for the fields in the  Pomeron trajectory and another for the fields in the meson trajectory.

\subsection{Pomeron in Holographic QCD in the Veneziano limit}
\label{sec:pomeron_eom}

The Pomeron is dual to the graviton Regge trajectory~\cite{Brower:2006ea}. To derive the equation of motion of the spin $J$ fields in the graviton's Regge trajectory we follow the same approach as in~\cite{Ballon-Bayona:2015wra, Ballon-Bayona:2017vlm}, that is we derive first the equation of motion of the graviton and then generalize it based on the two requirements mentioned above. 
Only the gluon part of the action $S_g$, contributes to the fluctuations in the spin-2 sector. Therefore the equation of motion of the $\mathrm{TT}$ components of the graviton in this class of models is, in the Einstein frame, the same as in IHQCD~\cite{Gursoy:2007er}
\begin{equation}
\nabla^2 h^{\mathrm{TT}}_{\mu \nu} + 2 {\dot{A}}^2 e^{-2A} h^{\mathrm{TT}}_{\mu \nu} = 0 \, ,
\label{eq:graviton_eom}
\end{equation}
which also equals equation (A.108) of~\cite{Arean:2013tja} for the $2^{++}$ glueballs. This equation reduces to (\ref{eq:spin_J_eom_ads}) for the AdS case since $\Delta = 4$ and $J  = 2$ for the AdS graviton. The corresponding equation of motion in the string frame can be obtained by noting that the warp factor in the Einstein frame  is related with the warp factor in the string frame $A_S$ through $A = A_S - 2 \Phi / 3$, and that the expected relationship between the string frame $\mathrm{TT}$ perturbations $h^{\mathrm{TT} \, S}_{\alpha \beta}$ and Einstein frame $\mathrm{TT}$ perturbations $h^{\mathrm{TT}}_{\alpha \beta}$ is $h_{\alpha \beta} = e^{- 4 \Phi / 3} h^S_{\alpha \beta}$. Then the equation of motion in the string frame is (dropping the superscript $S$ in 
$h^{\mathrm{TT} \, S}_{\alpha \beta}$)
\begin{equation}
\left[\nabla^2 - 2 e^{-2 A_s} \dot{\Phi} \nabla_z + 2 {\dot{A}_s}^2 e^{-2A_s} \right] h^{\mathrm{TT}}_{\alpha\beta} = 0 \, .
\label{eq:graviton_eom_string_frame}
\end{equation}

We propose that the equation of motion of the spin $J$ fields in the graviton's Regge trajectory is
\begin{equation}
\left[ \nabla^2 - 2 e^{-2A_s} \dot{\Phi} \nabla_z - \frac{\Delta ( \Delta - 4 )}{L^2} + J {\dot{A}_s}^2 e^{-2A_s} + e_g \left( J - 2\right) e^{-2A_s}  {\dot{\tau}}^2 \right] h^{\mathrm{TT}}_{\alpha_1 \dots \alpha_J} = 0 
\label{eq:spin_J_proposed} \, ,  
\end{equation}
where $e_g$ is a constant that will be fixed later by 
setting the soft pomeron intercept to $1.08$. We note that: $i)$ for $J = 2$ we get the the graviton's equation (\ref{eq:graviton_eom_string_frame}); $ii)$ for the conformal case, i.e. $A = - \log\left(z/L\right)$ and $\Phi$ and $\tau$ constant,   the equation reduces to (\ref{eq:spin_J_eom_ads}); $iii)$ the second term comes from the tree level coupling of a closed string, as appropriate for the graviton Regge trajectory in a large N approximation; $iv)$ following an effective field theory rational we could have included other terms proportional to derivatives of $A_s$, $\Phi$ and $\tau$, that is terms proportional to
\begin{align}
&e^{-2A_s} \left( \ddot{A}_s - \dot{A}_s^2 \right)\,, \quad e^{-2A_s} \dot{\Phi}^2 \, , \quad  \quad e^{-2A_s} \ddot{\Phi}  \, , \quad e^{-2A_s} \dot{A}_s \dot{\Phi} \, .
\label{eq:neglected_terms}
\end{align}
All these terms, of dimension inverse squared length, are compatible with the constraint $i)$ and also with constraint $ii)$ provided they multiply $J-2$.
Like the term proportional to $\dot{\tau}^2$, these terms are all subleading in the UV. However in the IR, where the wavefunctions of the associated Schr\"odinger problem are localised, the term $\dot{\tau}^2$ dominates and for this reason we will only consider this term.
We have also not included the terms $\ddot{\tau}$, $\dot{\tau} \dot{A}_s$ and $\dot{\tau} \dot{\Phi}$ because the background is symmetric under $\tau \rightarrow - \tau$.

The third term in (\ref{eq:spin_J_proposed}) is a mass term
obtained by the analytic continuation of the dimension of the exchanged operators $\Delta = \Delta\left(J\right)$. We shall set 
\begin{equation}
\frac{\Delta ( \Delta - 4 )}{L^2}  =    \frac{J^2 -4}{\lambda^{4/3}} \label{eq:delta_J_proposed_curve} \, .
\end{equation}
In the boundary theory the dimension of the operator $\mathcal{O}_J$ can be written as $\Delta = 2 + J + \gamma_J$, where $\gamma_J$ is the anomalous dimension. In free theory $\gamma_J = 0$.  The term we added in the right-hand side of (\ref{eq:delta_J_proposed_curve})
ensures the correct  UV asymptotic behaviour of   (\ref{eq:spin_J_proposed}) leading to the free theory result $h^{\mathrm{TT}}_{+\cdots+} \sim z^2$.
Beyond perturbation theory, the curve must pass through the point $J = 2$ and $\Delta = 4$ as it represents the energy-momentum tensor which is protected.
Equation (\ref{eq:delta_J_proposed_curve}) guarantees such properties.

The propagator for the spin $J$ fields in the graviton's Regge trajectory is the solution of
\begin{equation}
\left( \mathcal{D} \Pi \right)_{a_1 \cdots a_J, b_1 \cdots b_J} \left(x, \bar{x}\right) = i e^{2 \Phi} g_{a_1(b_1} \cdots g_{|a_J|b_J)} \delta_5 (x, \bar{x}) - \mathrm{traces} \, ,
\end{equation}
where the notation $g_{a_1(b_1} \cdots g_{|a_J|b_J)}$ means that symmetrisation
is applied only among the indices $b_i\, , i = 1 ,\cdots J$. As we have seen in the previous section, in the Regge limit we are only interested in the component $\Pi_{+\cdots+, - \cdots -}$. By using the identity (\ref{eq:gJ_t_def}) one can show that
\begin{equation}
\left[ \Delta_3 - e^{-2A_s} \left( 2 \dot{\Phi} \partial_z + 2 {\dot{A_s}}^2 + \ddot{A}_s - 2 \dot{A}_2 \dot{\Phi}  \right) - m^2_J (z) \right] G_J \left( z, \bar{z}, l_\perp \right) = - e^{2 \Phi } \delta_3 (y, \bar{y}) \, ,
\label{eq:gj_transverse}
\end{equation}
where  $l_\perp=x_\perp-\bar{x}_\perp$ and
$y = (z, x_\perp )$ and $\bar{y} = ( \bar{z}, \bar{x}_\perp )$ are points in the scattering transverse space with metric $ds^2_3 = e^{2A_s} \left[ dz^2 + d x^2_\perp\right] $.
$\Delta_3$ is the corresponding Laplacian and $m^2(z)$ is given by
\begin{align}
m_J^2(z) = & \left(J - 2 \right) \left[ \frac{J+2}{\lambda^\frac{4}{3}} + e_g \, e^{-2 A_s} {\dot{\tau}}^2  \right]  .
\end{align}
The homogeneous version of equation (\ref{eq:gj_transverse}) can be transformed in a Schr\"odinger problem through the Ansatz
\begin{equation}
G_J (z, \bar{z}, t) = e^{\Phi - \frac{A_s}{2}} \psi (z) \, ,
\end{equation}
where $\psi$ satisfies the Schr\"odinger equation
\begin{equation}
\left[ -\partial_z^2 -t + V_J (z) \right] \psi (z) = 0 \, ,
\end{equation}
with   $t = - q_\perp^2$ and 
\begin{equation}
V_J(z) = \frac{3}{2}\left( \ddot{A}_s - \frac{2}{3} \ddot{\Phi} \right) + \frac{9}{4} {\left( \dot{A}_s - \frac{2}{3} \dot{\Phi} \right)}^2 + e^{2 A_s} m_J^2 (z) \, .
\label{eq:schrodinger_potential_gluon}
\end{equation}
The spectrum for each integer $J$ discretises $t = t_n (J)$ and the corresponding eigenfunctions satisfy the identity $\sum_n \psi_n (z) \psi^{*}_n (\bar{z}) = \delta ( z - \bar{z} )$. Hence, the solution to equation (\ref{eq:gj_transverse}) is given by
\begin{equation}
G_J(z, \bar{z}, t ) = e^{\Phi + \bar{\Phi} - \frac{A_s}{2} - \frac{\bar{A}_s}{2}} \sum_n \frac{\psi_n (J, z)\, \psi_n^{*} (J, \bar{z})}{t_n (J) - t} \, .
\label{eq:gj_pomeron}
\end{equation}
and the $B$ function in equations (\ref{eq:sigma_gp_hol_gens}), (\ref{eq:sigma_gg_hol_gens}), (\ref{eq:sigma_pp_hol_gens}) is $\Phi - A_s/2$ when considering exchange of reggeons in the Pomeron trajectory.
 
\subsection{Meson Trajectory}

\begin{figure}[!t]
  \center
  \includegraphics[scale = 0.35]{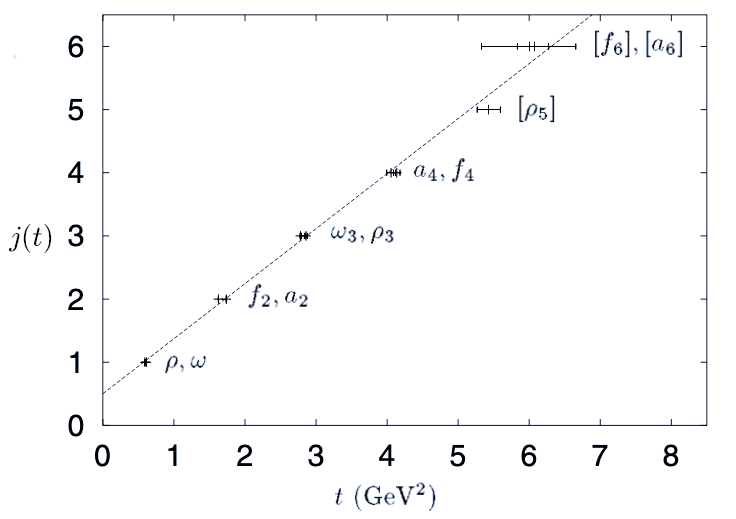} 
  \caption{Chew-Frautschi plot of four degenerate meson Regge trajectories. All the mesons shown are well established experimentally, except $\rho_5$, $f_6$ and $a_6$. The particle spins are plotted agains their squared masses $t$. Figure adapted from~\cite{Donnachie:2002en}.}
  \label{fig:meson_trajectories}
\end{figure}

Next we  discuss how to describe the dynamics of spin $J$ fields in the mesons trajectory. The Chew-Frautschi plot of figure~\ref{fig:meson_trajectories} shows two important properties of the mesons trajectories. The first is linearity which is usually used to extrapolate from the $t$-channel physical region to the $s$-channel scattering region where $t < 0$. That is, we find the best line to the points and use it to find $j(t)$ for $t < m_\rho^2$. In particular, for total cross-sections we are interested in the intercept value at $t = 0$. Another important property is the near degeneracy of the four trajectories $\{ f_2(1270), f_4(2050), \cdots \}$, $\{ a_2(1320), a_4(2040), \cdots \}$, $\{ \omega(780), \omega_3(1670), \cdots \}$ and $\{ \rho(770), \rho_3(1690), \cdots \}$.

Using the fact that the four trajectories are nearly degenerate and that we had a very good description of the spectrum for non-singlet and singlet vector mesons ($\rho$ and $\omega$) we will construct the holographic meson trajectory by generalizing the equation of motion of vector mesons to any spin $J$ field in the same trajectory. We will follow the same procedure of last section. Later we will validade our approach by showing how, with just one single parameter, one can simultaneously have a good description of scattering data, the spectrum of the meson trajectory, and an approximately linear Regge trajectory for the mesons.

From the action (\ref{eq:VM_new_action_string_frame}) for the vector field in the string frame, one gets the equation of motion
\begin{equation}
\nabla_{a} \left(e^{-\frac{10}{3} \Phi} \,V_f \,w_s^2 \,G \,\tilde{g}^{ac} \tilde{g}^{bd} F_{cd}  \right) = 0\,.
\label{eq:U1_eom_cov_der}
\end{equation}
In order to simplify the notation, from now on we assume all the warp factors and effective metrics are in the string frame.
As a first step to generalize   (\ref{eq:U1_eom_cov_der}) we write it as
\begin{align}
 \tilde{g}^{ab} \nabla_a \nabla_b A_\alpha & + \left( \frac{\partial_z \tilde{V}}{\tilde{V}G^2} +   \left( 3 \dot{A} \left(  \frac{1}{G^2} - 1\right) - \frac{\dot{G}}{G^3} \right)  \right) e^{-2 A}
 \big(\nabla_z A_\alpha - \nabla_\alpha A_z \big) - \notag \\
&- \left( 1 - \frac{1}{G^2} \right) \nabla_\alpha \nabla_\lambda A^\lambda +  \left( \frac{\ddot{A}}{G^2} + 3 \dot{A}^2  \right) e^{-2 A} A_\alpha = 0 \, ,
\end{align}
where $\tilde{V} = e^{-\frac{10}{3} \Phi} V_f w_s^2$. In the conformal limit (pure AdS with constant dilaton and tachyon) the above equation reduces to
\begin{equation}
\big( \nabla^2 - M^2 \big) A_\alpha = 0 \, ,
\end{equation}
with ${\left( L M \right)}^2= - 4$, as expected for a bulk field dual to the current operator $\bar{\psi} \gamma^\mu \psi$ with spin $J = 1$ and protected dimension $\Delta = 3$.

We now propose an equation of motion for the symmetric, traceless and transverse spin $J$ field $h_{a_1 \cdots a_J}$ in the meson Regge trajectory. 
As in the graviton case, we are interested in the $\mathrm{TT}$ part $h^{\mathrm{TT}}_{\alpha_1 \cdots \alpha_J}$ which are the propagating degrees of freedom that decouple from the other components after decomposing the field in $SO(1,3)$ irreducible representations.
We propose the equation
\begin{align}
& \left[ \tilde{g}^{ab} \nabla_a \nabla_b  + \left( \frac{\partial_z \tilde{V}}{\tilde{V} \, G^2} +  3 \dot{A} \left(  \frac{1}{G^2} - 1\right) - \frac{\dot{G}}{G^3} \right) e^{-2 A} \left( \partial_z - (J-1) \dot{A} \right)  -\right. \label{eq:spin_j_meson_eom} \\
&\left. \left( 1 - \frac{1}{G^2} \right)  \nabla_{\alpha_1} \nabla_\lambda -  \left(  \Delta \left( \Delta - 4 \right) \dot{A}^2 - J \frac{\ddot{A}}{G^2}  \right) e^{-2 A} + e_m \left(J-1\right) e^{-2A} {\dot{\tau}}^2 \right] h^{\mathrm{TT}}_{\alpha_1 \cdots \alpha_J} = 0 \,. \notag 
\end{align}
We note that: $i)$ this equation reduces to the vector meson equation of motion for $J = 1$; $ii)$ in the AdS case the second, third and fifth terms vanish, reducing this equation to the equation of motion of the  $\mathrm{TT}$ components of spin $J$ fields in AdS; $iii)$ Following the same logic as for the fields in the Pomeron trajectory we only included the 
two-derivative term in $\dot{\tau}^2$ that is compatible with  
 $i)$ and $ii)$. Also, as in the case of the Pomeron, the $\Delta(J)$ curve  follows from the analytic continuation of the dimension of the exchanged operators and imposing the correct UV asymptotic behaviour of the spin $J$ fields:

\begin{equation}
\Delta \left( \Delta - 4 \right) = - 3 + \frac{e^{2A}}{\dot{A}^2} \frac{J^2 - 1}{\lambda^{\frac{4}{3}}} \, .
\end{equation}
The last term guarantees the correct UV behaviour  of the spin $J$ fields given in (\ref{eq:asymptotics}),
while the first one ensures that the curve passes through the protected point $\Delta = 3$, $J = 1$.

Equation (\ref{eq:spin_j_meson_eom}) can be brought to Schr\"odinger form by first rewriting it in terms of the $u$ variable defined previously. After that we can write the spin $J$ field as
\begin{equation}
h^{\mathrm{TT}}_{\alpha_1 \cdots \alpha_J} = \epsilon_{\alpha_1 \cdots \alpha_J} e^{i q \cdot x} \frac{e^{\left(J-1\right)A}}{\sqrt{e^{-\frac{10}{3}\Phi}\,V_f\, w_s^2 \,e^A}}\, \psi (u ),
\end{equation}
where $\psi$ satisfies the Schr\"odinger equation
\begin{equation}
- \frac{d^2 \psi}{du^2} + V_J (u) \,\psi = t\, \psi.
\label{eq:schrodinger_problem_meson}
\end{equation}
The Schr\"odinger potential is given by
\begin{equation}
V_J(u) = V_V(u) + \left(J^2 - 1\right) e^{2 A - \frac{4}{3} \Phi} + (J-1)\! \left[G^2 \dot{A}^2 - \frac{\dot{A}\dot{G}G+G^2 \ddot{A}}{G^2} - e_m G^2 \dot{\tau}^2 \right]  ,\label{eq:meson_schrodinger_potential}
\end{equation}
where $V_V$ is the Schr\"odinger potential of the vector mesons given in (\ref{eq:V_V}). Here the dots mean derivatives with respect to the u variable.
We will denote the eigenvalues of this Schr\"odinger potential as $t_n(J)$ and the corresponding eigenfunctions as $\psi_n(J, u)$.

The propagator for this spin $J$ field obeys an equation of the type
\begin{equation}
\left(\mathcal{D} \Pi\right)_{a_1 \cdots a_J, b_1 \cdots b_J} = i g_{a_1(b_1} \cdots g_{|a_J|b_J)} \delta_5 (x, \bar{x})\, \frac{G}{e^{-\frac{10}{3}\Phi}V_f w_s^2} \,,
\end{equation}
where $\mathcal{D}$ is a differential operator defined by  (\ref{eq:spin_j_meson_eom}).
In the Regge limit we will be interested in the components $\Pi_{+\cdots+,-\cdots-}$ and for this particular case
\begin{equation}
\mathcal{D} \Pi_{+\cdots+,-\cdots-} = i {\left( - \frac{e^{2A}}{2}\right)}^J  \delta_5 (x, \bar{x}) \,\frac{G}{e^{-\frac{10}{3}\Phi}V_f w_s^2}\,.
\end{equation}
Consider now the integral
\begin{equation}
\int \frac{dw^+ dw^-}{2} \, \Pi_{+\cdots+, - \cdots -} = -i {\left(-\frac{1}{2}\right)}^J e^{\left(J-1\right)(A+\bar{A})} G_J(u, \bar{u}, l_\perp)\,,
\end{equation}
that defines the transverse propagator $G_J$.
Applying the operator $\mathcal{D}$ on both sides of this equation and noting that 
\begin{align}
&\mathcal{D} \left[ e^{\left(J-1\right) \left(A+\bar{A}\right)} G_J (u, \bar{u}, l_\perp) \right] = e^{\left(J-1\right)\left(A+\bar{A}\right)} \mathcal{D}_3 G_J (u, \bar{u}, l_\perp) \, ,
\end{align}
where $\mathcal{D}_3$ is the differential operator
\begin{align}
&\mathcal{D}_3 = e^{-2A} \partial_u^2 + e^{-2A} \left( \dot{A} - \frac{10}{3} \dot{\Phi} + 2 \frac{w_s' \dot{\Phi}}{w_s} + \frac{\partial_u V_f}{V_f} \right) \partial_u + e^{-2A} \partial^2_{l_\perp} +  \\
& + \left(J-1\right)  e_m e^{-2A} G^2 {\dot{\tau}}^2 - \frac{J^2-1}{\lambda^{4/3}} - \left(J-1\right) e^{-2A} \left[G^2 \dot{A}^2 - \frac{\dot{A}\dot{G}G+G^2 \ddot{A}}{G^2} \right]  \, ,
\notag
\end{align}
we conclude that 
$G_J$ satisfies 
\begin{equation}
\mathcal{D}_3\, G_J (u, \bar{u}, l_\perp) = - \delta_3 (y, \bar{y})  \,\frac{G}{e^{-\frac{10}{3} \Phi} V_f w_s^2}
\label{eq:transverse_propagator}
\end{equation}
where $y$ and  $\bar{y}$ are coordinates in transverse space, as defined after (\ref{eq:gj_transverse}).

To solve  (\ref{eq:transverse_propagator}) we consider its homogeneous version and the following Ansatz
\begin{equation}
G_J(u, l_\perp) = \frac{e^{i q \cdot l_\perp}}{\sqrt{e^{-\frac{10}{3}\Phi}V_f w_s^2 e^{A}}}\, \psi(u)\,.
\end{equation}
It follows that $\psi$ is a solution of the Schr\"odinger problem (\ref{eq:schrodinger_problem_meson}).
Since the eigenfunctions of the Schr\"odinger potential satisfy $\sum_n \psi_n (u) {\psi_n (\bar{u})}^* = \delta(u - \bar{u})$
the solution to   (\ref{eq:transverse_propagator}) is
\begin{equation}
G_J (u, \bar{u}, t) =  \left.\left(e^{-\frac{10}{3}\Phi}V_f w_s^2 e^A \right)^{-1/2}\right|_u  \left.\left(e^{-\frac{10}{3}\Phi}V_f w_s^2 e^A \right)^{-1/2}\right|_{\bar{u}}  
\sum_n \frac{\psi_n(u) \psi_n (\bar{u})}{t_n(J) - t} \, .
\end{equation}
Thus, the implications of this result for   (\ref{eq:sigma_gp_hol_gens}), (\ref{eq:sigma_gg_hol_gens}) and (\ref{eq:sigma_pp_hol_gens}) is that 
\begin{equation}
B = - \frac{1}{2} \log\left( e^{-\frac{10}{3}\Phi}V_f w_s^2 e^A \right) \, ,
\end{equation}
when considering exchange of reggeons in the meson trajectory.

\section{Fit of $\gamma \gamma $, $\gamma p$ and $pp$ total cross-sections in holographic model}
\label{seq:xseq_fits}
In this section we will test the presented phenomenological model for $\gamma \gamma $, $\gamma p$ and $pp$ total cross-sections against the hadronic cross-section data files from the Particle Data Group~\cite{Zyla:2020zbs}. These data sets are formed from experimental results obtained by different groups over the last decades. The datasets of  $\sigma(\gamma p \rightarrow X)$ and $\sigma(p p \rightarrow X)$ have cross-section values as a function of the laboratory momentum of an incoming on-shell photon or proton, respectively. 
A calculation of the respective center of mass energy $\sqrt{s}$ was performed before starting the fits. We also considered only subsets of data with $\sqrt{s} > 4 \, {\mathrm{GeV}}$ for $\sigma(\gamma \gamma \rightarrow X)$, $\sigma(\gamma p \rightarrow X)$ and $\sigma(p p \rightarrow X)$,  yielding 39, 45 and 115 experimental points, respectively. 

We find the best set of parameter values $\alpha_i$ by minimising the $\chi^2$ quantity
\begin{align}
\chi^2 = \sum_{n = 1}^{N} {\left( \frac{O^{{\mathrm{pred}}}_k(\alpha_i) - O^{{\mathrm{exp}}}_k}{\sigma_k} \right)}^2 \, ,
\label{eq:chi2_def}
\end{align}
that is, the sum of the weighted difference squared between experimental data and model predicted values where the weight is the inverse of the experimental uncertainty. In our fits the parameters $\alpha_i$ are the couplings $k^{g/m}_{j_n}$ and $\bar{k}^{g/m}_{j_n}$ defined in equations (\ref{eq:gp_couplings_def}), (\ref{eq:gg_couplings_def}) and (\ref{eq:pp_couplings_def}), where the superscript refers the coupling to the pomeron or meson trajectory. Usually a fit is deemed of good quality if the quantity $\chi^2_{\mathrm{d.o.f.}} \equiv \chi^2 / (N - N_\mathrm{par}) \sim 1$, where $N_\mathrm{par}$ is the number of parameters $\alpha_i$ to be fitted. 
In  (\ref{eq:chi2_def}) $O_k$ represents a generic data point of one or several observables mentioned previously and, usually,  $\sigma_k$ is the experimental uncertainty associated with the measurement.
Some data points have uncertainties in the values of $s$ (e.g. in $\gamma \gamma \rightarrow X$ that is always the case because it is  measured experimentally). To account for this we calculate the total cross-section for $s + \Delta s$ and $s - \Delta s$, and evaluate
\begin{align}
\sigma_{\mathrm{eff.}} = {\mathrm{max}} &\left(|\sigma^{\mathrm{pred.}}\left(s + \Delta s\right)-\sigma^{\mathrm{pred.}}\left(s\right)|\,,  |\sigma^{\mathrm{pred.}}\left(s - \Delta s\right)-\sigma^{\mathrm{pred.}}\left(s\right)| \right).
\end{align}
For these cases  $\sigma_k = \sqrt{{\left(\sigma_{\mathrm{exp.}}\right)}^2+{\left( \sigma_{\mathrm{eff.}}\right)} ^2}$ where $\sigma_{\mathrm{exp.}}$ is the experimental error.

\begin{figure}[!t]
  \center
  \includegraphics[scale = 0.7]{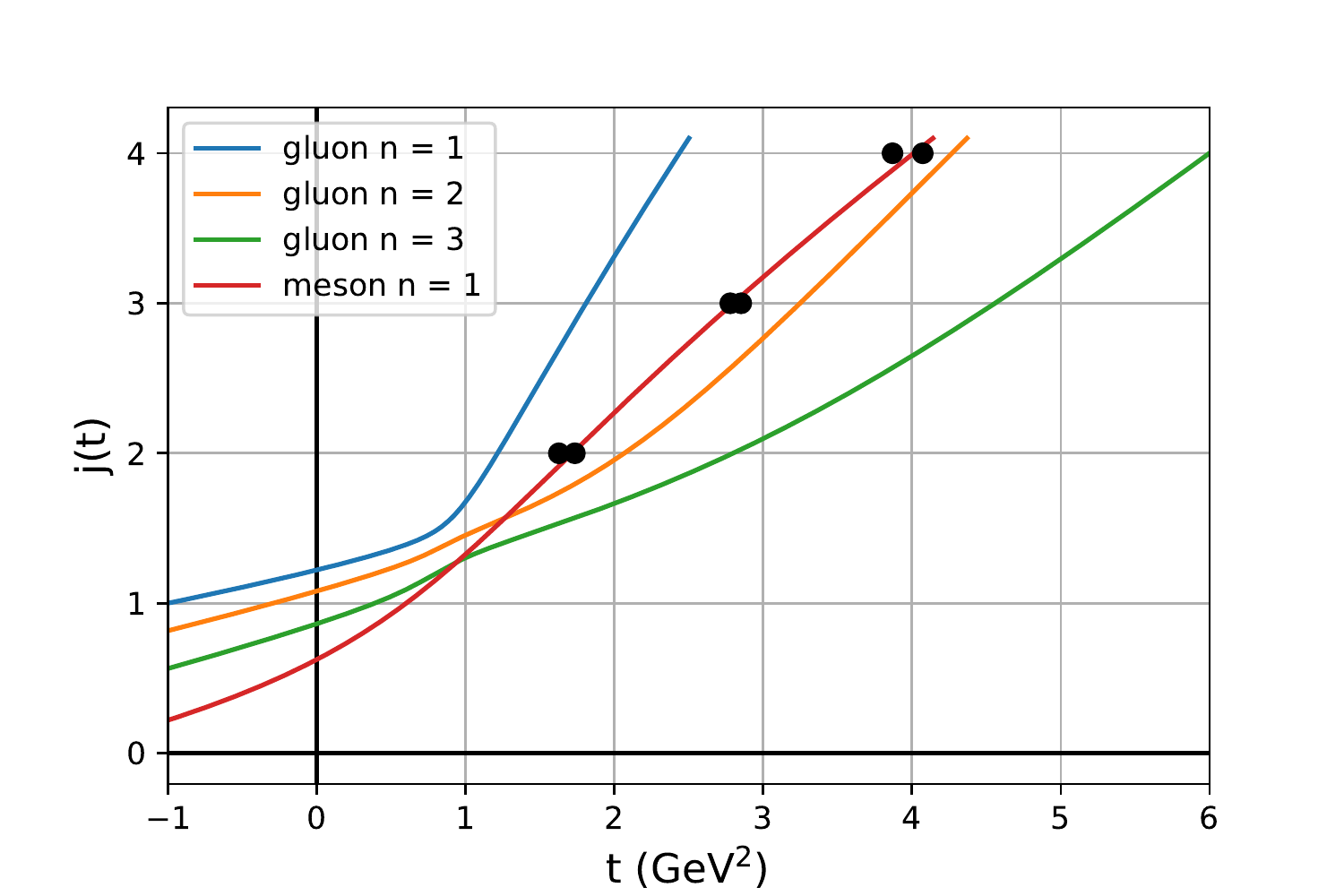} 
  \caption{The first three Regge trajectories of the gluon kernel and the first meson trajectory of the meson kernel used in the cross section fits. They result from solving the Schr\"odinger problems (\ref{eq:schrodinger_potential_gluon}) and (\ref{eq:meson_schrodinger_potential}) for several values of $J$. The black dots are the experimental values of the mesons $f_2$, $a_2$, $\rho_3$, $\omega_3$, $f_4$ and $a_4$.}
  \label{fig:regge_trajectories}
\end{figure}

By minimising (\ref{eq:chi2_def}) with all the data mentioned above we can find the best values for the potentials parameters $e_g$ and $e_m$, as well as for the coupling values $k_{j_n}$ and $\bar{k}_{j_n}$ with $ n = 1, 2, \cdots$ for each trajectory. We will follow another approach. As shown in~\cite{Ballon-Bayona:2017vlm}, the first two trajectories of the gluon kernel can be identified to the hard and soft pomeron trajectories. Here we will fix $e_g$ by demanding that the intercept of the soft pomeron trajectory is $1.08$. We also identify the meson trajectory with the first trajectory of the meson kernel and we will fix $e_m$ such that we have agreement with the meson masses of $(f_2, a_2)$, $(\rho_3, \omega_3)$ and $(f_4, a_4)$ for $J = 2, 3, 4$, respectively. The numerical values of these parameters were found to be $0.246$ and $1.712$ for $e_{g}$ and $e_{m}$, respectively. The intercept values obtained from these parameter values can be found in table~\ref{table:regge_traj_intercepts} and the leading trajectories are shown in figure~\ref{fig:regge_trajectories}. The corresponding masses of the mesons for $J = 2, 3, 4$ are present in table~\ref{table:higher_spin_meson_masses}. The intercept of the hard pomeron is close to $1.17$ which was the one found in the IHQCD model~\cite{Ballon-Bayona:2017vlm, Amorim:2018yod}. We also note that the intercept of the 4th Pomeron trajectory and of the meson trajectory are very close to 0.55 found in the hadronic cross-section fits of~\cite{Donnachie:2002en}.
\begin{table}[t!]
\centering
\caption{Values of the intercepts of the first four trajectories of the Pomeron kernel and of the first two trajectories of the meson kernel. These values were obtained with $e_g = 0.246$ and $e_m = 1.712$. }
\begin{tabular}{|c|c|}
\hline
Intercept & Intercept value  \\
\hline\hline
$j^{\mathrm{g}}_1$ & $1.22$  \\
\hline
$j^{\mathrm{g}}_2$ & $1.08$\\
\hline
$j^{\mathrm{g}}_3$ & $0.862$ \\
\hline
$j^{\mathrm{g}}_4$ & $0.574$ \\
\hline\hline
$j^{\mathrm{m}}_1$ & $0.625$ \\
\hline
$j^{\mathrm{m}}_2$ & $0.246$ \\
\hline
\end{tabular}
\label{table:regge_traj_intercepts}
\end{table}
\begin{table}[t!]
\centering
\caption{Meson masses in GeV for $J=2,3,4$ obtained for $e_m = 1.712$ and the corresponding experimental values.}
\vspace{0.5cm}
\begin{tabular}{|c|c|c|}
\hline
meson  &  mass & predicted mass  \\
\hline\hline
$f_2(1270)$ / $a_2(1320)$ & 1.2755 / 1.3169 & 1.310 \\
\hline
$\rho_3(1690)$ / $\omega_3(1670)$ & 1.6888 / 1.667 & 1.673 \\
\hline
$f_4(2050)$ / $a_4(2040)$ & 2.018 / 1.967 & 2.003 \\
\hline
\end{tabular}
\label{table:higher_spin_meson_masses}
\end{table}

\begin{figure}[!t]
  \center
  \includegraphics[scale = 0.6]{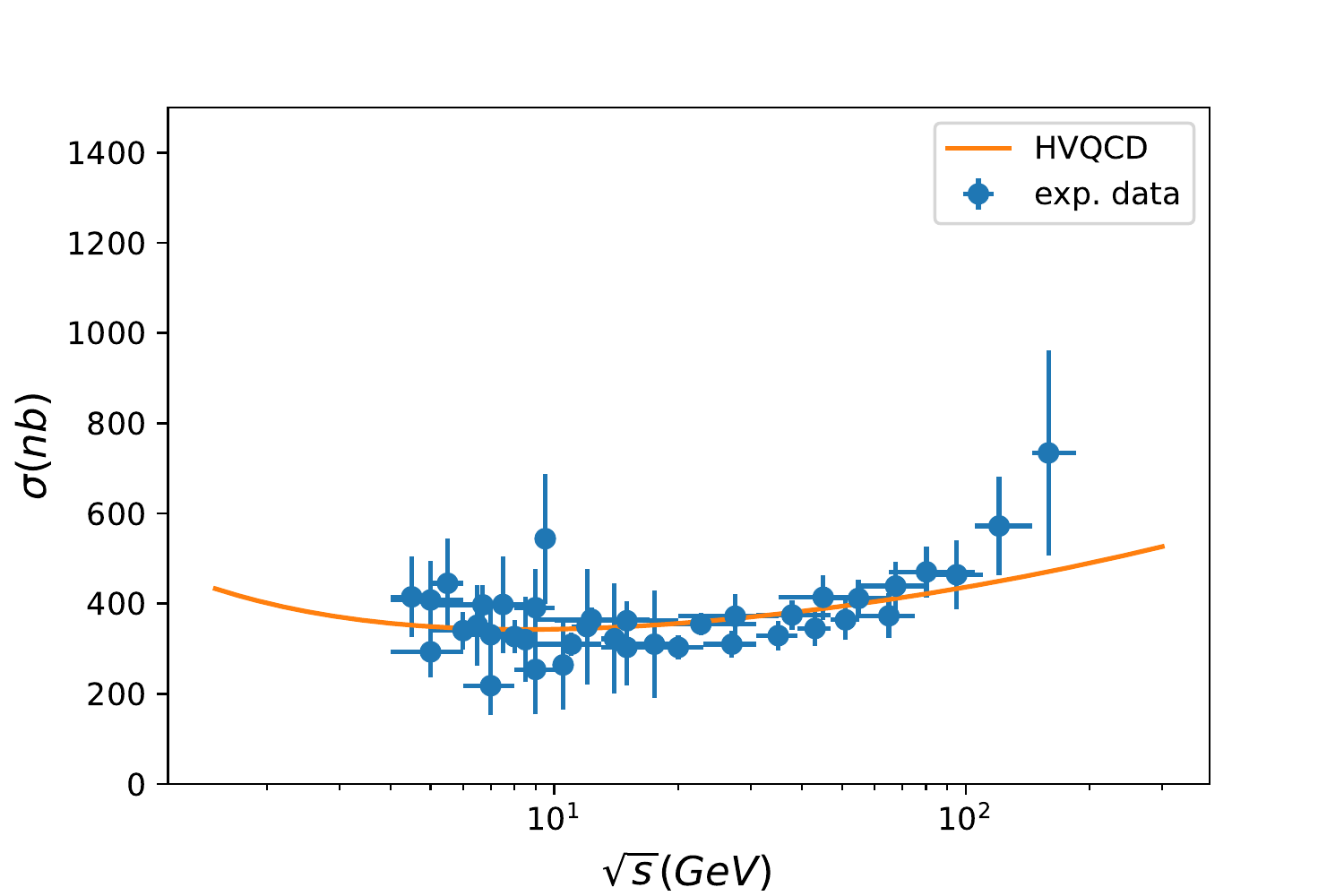} 
  \caption{Fit of $\sigma ( \gamma \gamma \rightarrow X )$ vs experimental points. The curve was obtained using the values from table~\ref{table:best_fit_pars_all_sigma}.}
  \label{fig:sigma_gg}
\end{figure}

\begin{figure}[!t]
  \center
  \includegraphics[scale = 0.6]{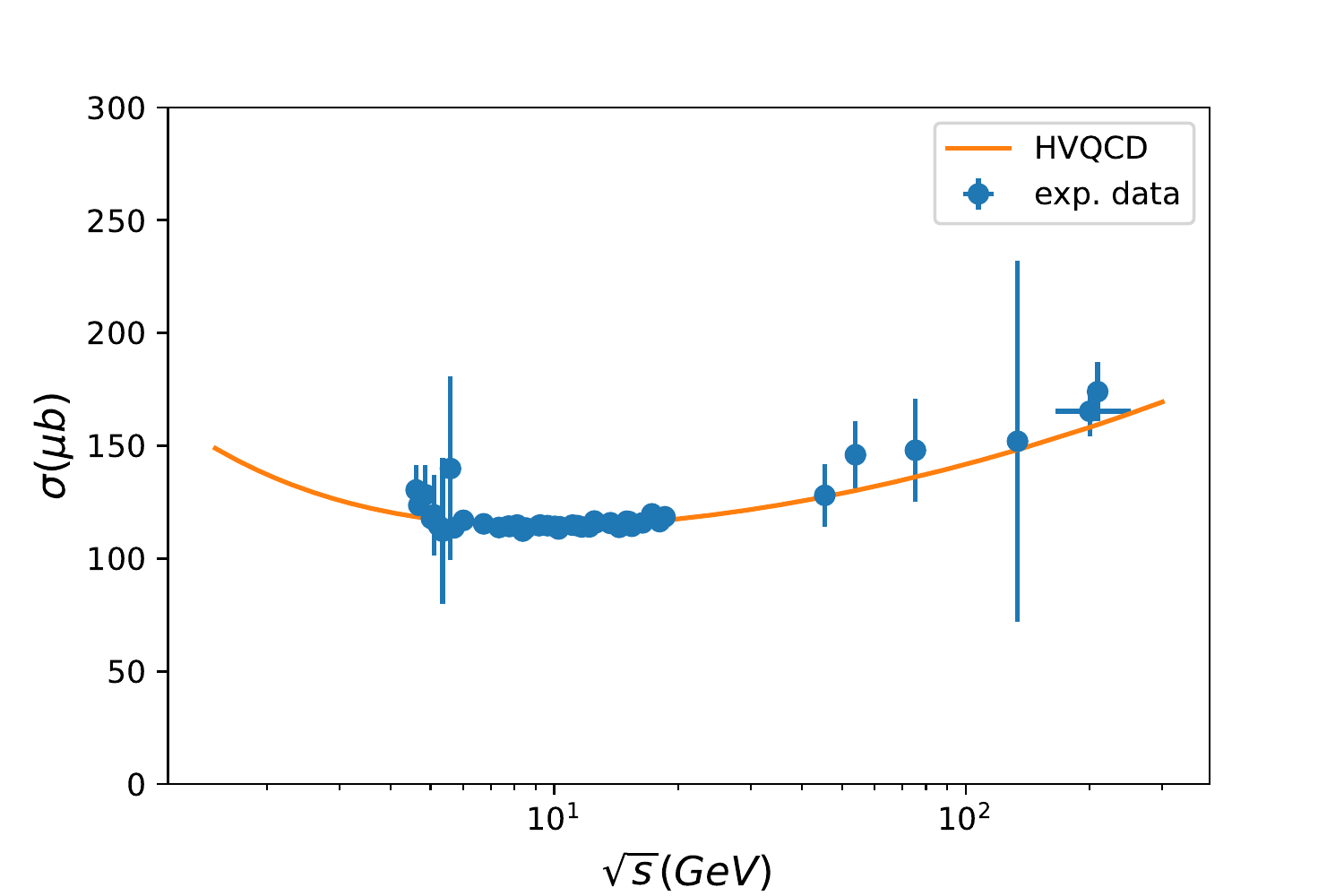} 
  \caption{Fit of $\sigma ( \gamma p \rightarrow X)$ vs experimental points. The curve was obtained using the values from table~\ref{table:best_fit_pars_all_sigma}.}
  \label{fig:sigma_gp}
\end{figure}

\begin{figure}[!t]
  \center
  \includegraphics[scale = 0.6]{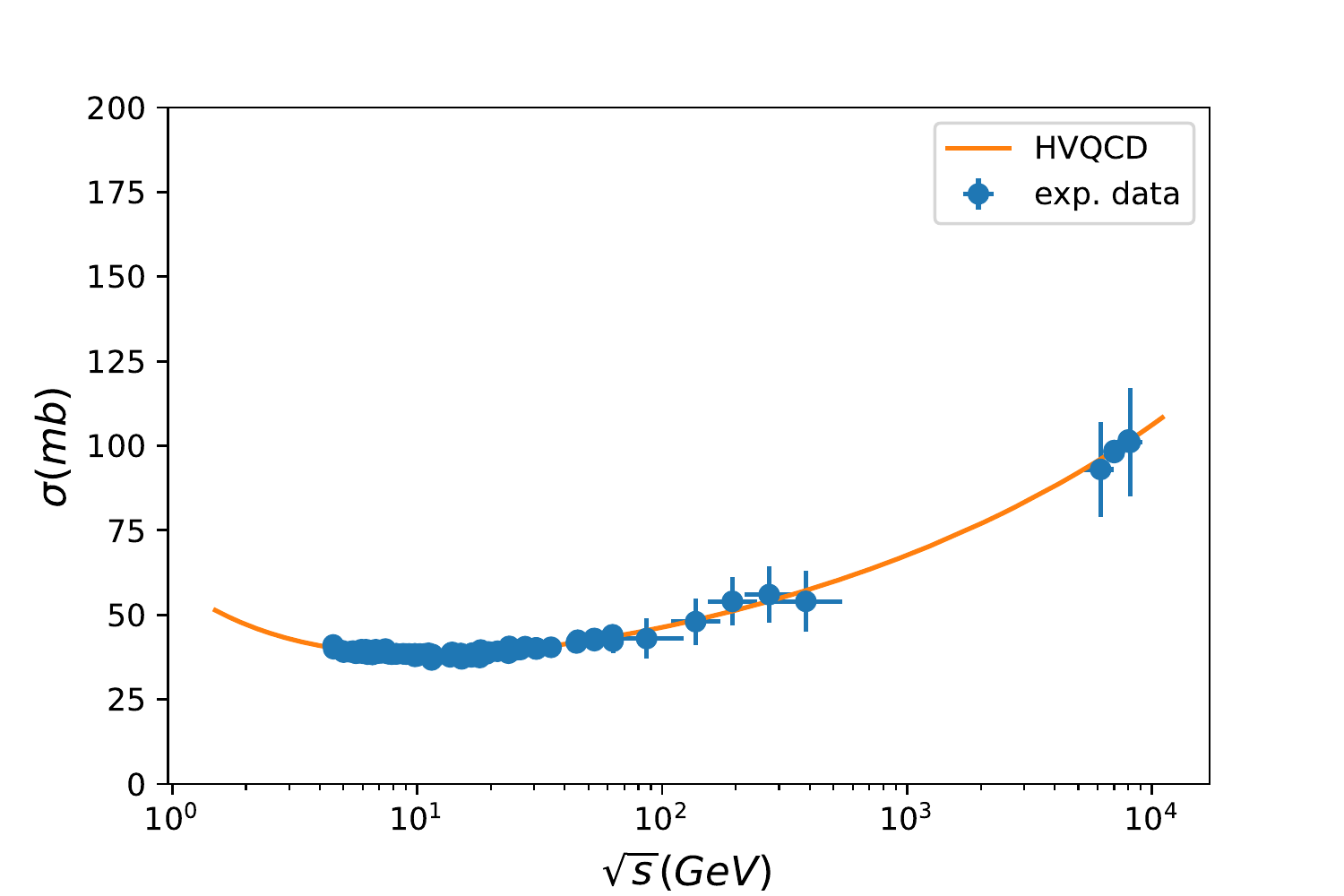} 
  \caption{Fit of $\sigma ( p p \rightarrow X)$ vs experimental points. The curve was obtained using the values from table~\ref{table:best_fit_pars_all_sigma}.}
  \label{fig:sigma_pp}
\end{figure}

After fixing the kernel parameters $e_g$ and $e_m$, we fit the total cross-section data by minimising~\eqref{eq:chi2_def} with respect to the couplings $k^{g/m}_{j_n}$ and $\bar{k}^{g/m}_{j_n}$. Considering only the first three Pomeron trajectories and the first meson trajectory with the intercepts of table~\ref{table:regge_traj_intercepts} we obtain a $\chi^2_{\mathrm{d.o.f.}}$ of $0.74$ for a total of 199 experimental points and 8 parameters. The best fit parameters are present in table~\ref{table:best_fit_pars_all_sigma} and the comparison of the experimental data with the predictions of the model for these parameters is present in figures \ref{fig:sigma_gg}, \ref{fig:sigma_gp} and \ref{fig:sigma_pp}. These results show that one does not need to assume a linear meson trajectory after fixing it at $t > 0$ in order to describe total cross-section data, as it is often assumed in the literature, leading to an the intercept value of 0.55. The Regge trajectory is convex  line, which  yields a slightly higher intercept.  
\begin{table}[t!]
\centering
\caption{Values of the couplings for the joint fit of $\sigma\left(\gamma \gamma \to X\right)$, $\sigma\left(\gamma p \to X\right)$ and $\sigma\left(p p \to X\right)$ data. There are 199 experimental points giving a  $\chi^2_{d.o.f.}=0.74$.}
\vspace{0.5cm}
\begin{tabular}{|c|c|c|c|}
\hline
$\gamma$ coupling & value & proton coupling & value  \\
\hline\hline
$k^g_{j_1}$ & 0.0634496 & $\bar{k}^g_{j_1}$ & 2.23798 \\
\hline
$k^g_{j_2}$ &-0.119512 & $\bar{k}^g_{j_2}$ & -15.485 \\
\hline
$k^g_{j_3}$ &0.0180875 & $\bar{k}^g_{j_3}$ & 7.11293 \\
\hline\hline
$k^m_{j_1}$ & 0.043448 & $\bar{k}^m_{j_1}$ & 12.8753  \\
\hline
\end{tabular}
\label{table:best_fit_pars_all_sigma}
\end{table}

\section{Conclusions}
\label{seq:conclusions}

In this article, we studied Regge theory in a full-fledged holographic model (V-QCD), which includes backreaction of quark degrees of freedom to the gluon dynamics in QCD. 
The main new results can be divided into two categories: Firstly, we made progress with the comparison of the model with QCD data by carrying out a detailed fit of the model paratemeters to the meson spectrum. Secondly, we developed a scheme to describe higher spin mesons and Regge trajectories in this model, and applied it to analyse the total QCD cross sections of scattering precosses having protons and photons in the inital state.

As explained in section~\ref{seq:hvqcd_sum}, the holographic model is strongly constrained by the requirement that it agrees with known features of QCD such as confinement, chiral symmetry breaking, asymptotic linearity of meson trajectories, qualitatively correct dependence of the spectrum on the quark mass, correct response to small chemical potential at small temperatures, and asymptotic freedom with correct dimensions (and anomalous dimensions) of the most important operators of QCD at weak coupling. Most of the remaining parameters, which are not determined by such qualitative considerations, amount to tuning of the various potentials of the V-QCD action at intermediate values of the coupling. In this article we have chosen to tune these parameters such that the meson spectrum of the model agrees well with experimental QCD data. 
The number of fitted parameters is large, because we want to make sure that our Ansatz for the potentials covers essentially all of the parameter space left free by the constraints listed above. However since the effect of all these parameters on the potentials is relatively small, the dependence of the result for the meson masses on the parameters is weak. In other words, one obtains a rather good description of the QCD spectrum for any reasonable values of the parameters, and the task carried out in this article is to tune the masses to agree as well with experimental values as possible.

Because we were mostly interested in the Regge physics in this article, we chose a strategy where we only fitted the meson masses with spins $J=0$ and $J=1$, but did not consider other data such as decay constants or thermodynamic potentials. Notice also that we included radial excitations     
with high masses and these state were fitted with the same weight as the ``important'' low-lying states such as the pions and the $\rho$-meson. That is, the fit was tailored for the purpose of studying the Regge physics where reproducing the correct asymptotics of the trajectories is important. It is anyhow interesting that the results of the fit  are in good agreement with those obtained in~\cite{Jokela:2018ers} where essentially the same Ansatz was compared to the lattice data for the thermodynamics of Yang-Mills theory and QCD at finite temperature, and in rough agreement with the value of the parameter $c$ of~\cite{Gursoy:2016ofp,Gursoy:2020kjd} as we explained in section~\ref{sec:fit}.
We also remark that the fit carried out here appears to be more constraining than those carried out in the earlier references: our fit favors $W_0 \approx 2.5$ whereas this parameter was left unconstrained by the comparison to lattice thermodynamics.
We obtained a very good fit for the spin 1 and pseudoscalar meson masses. This fit was  more extensive than that carried out in the probe limit in the closely related model of~\cite{Iatrakis:2010zf,Iatrakis:2010jb}. It also compares favorably to work in simpler holographic models and in models inspired by gauge/gravity duality, such as the hard~\cite{Erlich:2005qh,DaRold:2005mxj} and soft wall models~\cite{Karch:2006pv}, light front holography~\cite{deTeramond:2005su,Brodsky:2014yha}, and the holography inspired stringy hardon model~\cite{Sonnenschein:2014jwa,Sonnenschein:2018fph}.

There are several ways to further develop the holographic model and the fitting procedure in the future. A simple project would be to redo the fit to meson spectrum with the aim of producing a model for all purposes, which would mean to weight more the mesons having low masses and also consider other experimental data relevant for the zero temperature vacuum such as decay constants, the flavor singlet scalar and pseudoscalar states, topological susceptibility, the $S$-parameter and so on. The potentially challenging issue, which we noticed while doing the fit of this article, is that the scalar meson masses agree poorly with the experimental values. This issue  would need to be solved. A more ambitious project would be to carry out a simulatenous comparison of the zero temperature and finite temperature data, i.e. to also include the data for lattice thermodynamics at finite temperature and potentially also at finite magnetic field. Good agreement of the model parameters obtained by fitting the zero and finite temperature data independently suggests that such a project is feasible. As a part of the project, it might make sense to also generalise the model to include flavor dependent quark masses and flavor dependent coupling of the magnetic field to the quarks.

Having fixed 
the action for the geometry and the actions for scalar and vector mesons through the fit,   
we  proposed the  dynamics for the spin $J$ fields dual to the gluon and quark twist two operators. This dynamics is controlled by two parameters that were fixed by making the second Pomeron trajectory intercept to be $1.08$ (the known soft-pomeron intercept) and by reproducing quite accurately  the masses of the mesons with $J = 2, 3, 4$. We have found that the hard pomeron intercept is close to 1.17 in IHQCD. We also note that the fourth Pomeron trajectory and the leading meson trajectory that we obtained have  intercepts close to the meson intercept value of 0.55 commonly found in the literature. Using the first three pomeron trajectories and the first meson trajectory we had a very good fit of the total cross-sections of $\gamma \gamma$, $\gamma p$ and $pp$ scattering. We note here that although our meson intercept is close to 0.55, the corresponding trajectory is non-linear in the $t>0$ region (see figure \ref{fig:regge_trajectories}), as it is often assumed. This suggests that the linearity of the trajectory might be at best a very good approximation to obtain an intercept that explains total cross-section data, and not its true shape in this region.

This work can now be extended to other processes like the proton structure functions $F_2^{\mathrm{p}}$ and 
$F_L^{\mathrm{p}}$  
and the photon structure function $F_2^{\mathrm{\gamma}}$, as done in~\cite{Amorim:2021ffr}.  In this model their holographic expressions are given respectively by equations (\ref{eq:hol_f2}), (\ref{eq:hol_fl}) and (\ref{eq:hol_f2_gamma}) from the appendix~\ref{appendix:pgammastr}.  
From these expressions one can see that: $i)$ the Bjorken $x$ behaviour depends solely on the value of the intercept; and $ii)$ the $Q^2$ dependence comes from an integral involving the non-normalizable mode of the $U(1)$ 
gauge field $f_Q$,  
functions of the background fields and the 
normalisable wave functions $\psi_n$  
of the pomeron and meson kernels. Having the background fixed, the first two classes of functions are uniquely determined while the $\psi_n$'s might be controlled by phenomenological parameters 
(similar to $e_g$ and $e_f$)  
of the equations of motion of the spin $J$ fields. In this work, with the values of $e_g$ and $e_f$ fixed by the spectrum of higher spin mesons, we may have obtained the correct Bjorken $x$ behaviour, but it is not guaranteed that the wave functions yield the observed dependence of the structure functions on $Q^2$. Hence, one needs to consider the neglected terms of equation (\ref{eq:neglected_terms}) in order to get good agreement between our model and data. Of course this involves adding more parameters to these fits, making it harder to find a minimum for the $\chi^2$ function that describes satisfactory all the data being considered. If successful, this would also extend~\cite{Amorim:2021ffr} by including also $pp$ total cross-section data in a consistent holographic Regge analysis.

Since this model reproduces well the masses of the towers of $\rho$ and $\pi$ mesons, one could also test this approach against the Vector Meson Production (VMP) data from HERA with a $\rho$ meson in the final state and to include $\pi^0 p$ total cross-section data. This will introduce extra parameters in this holographic Regge model. In the $\pi^0 p$ total cross-section we could introduce couplings between the pions and the spin $J$ fields of the different trajectories. On the other hand, the VMP data would promote the coupling constants $k_J$ to functions of $t$ since in differential cross-section data we work with more than one value of $t$. In order to control the amount of parameters to be introduced in this model, it would be interesting to fix the couplings, or the functional form of $k_{j_n(t)}$, to other observables where these couplings might be important. This could be done by first using a theoretical well motivated ansatz for $k_{j_n(t)}$ with some free parameters. These free parameters could then be fixed by the experimental values of decay rates. As an example of this, $k^m_{j_1(m^2_{f_2})}$ and $\Gamma(f_2 \to \gamma \gamma)$ are related through equation (\ref{eq:gauge_field_spin_J_coupling}) for the $J = 2$ case.

To describe successfully VMP in this holographic setting we need a better approximation to the proton state than assuming it to be a scalar field. In this work this is sufficient, since the integrals that involve the proton wavefunction are absorbed in the fitting parameters. However, in the VMP case the proton state is in an integrand multiplying wavefunctions of the Pomeron and meson kernel that depend on $t$. Moreover, having 
a good model for  
the proton state could also allow to extend this work directly by including available data of $pp$ differential cross-section data. In holography, baryons are dual to solitons in the bulk and hence the problem is reduced to the calculations of these solutions~\cite{Witten:1998xy}. Such solitons have been studied in the literature in the Witten-Sakai-Sugimoto model~\cite{Hata:2007mb,Kim:2007zm} and in hard-wall models~\cite{Pomarol:2007kr,Pomarol:2008aa}. In the V-QCD model, baryons have so far only been considered by employing an approximation scheme~\cite{Ishii:2019gta}, and the construction of the soliton solutions is work in progress.  
Such a solution could be a good starting point to model the proton state in holographic Regge theory.

\section*{Acknowledgments}

This research received funding from the Simons Foundation grants 488637  (Simons collaboration on the Non-perturbative bootstrap). 
Centro de F\'\i sica do Porto is partially funded by Funda\c c\~ao para a Ci\^encia e a Tecnologia (FCT) under the grant
UID-04650-FCUP.
 AA is funded by FCT under the IDPASC doctorate programme with the fellowship  PD/BD/114158/2016.
The research of MJ was supported by an appointment to the JRG Program at the APCTP through the Science and Technology Promotion Fund and Lottery Fund of the Korean Government.
MJ was also supported by the Korean Local Governments -- Gyeongsangbuk-do Province and Pohang City.

\appendix

\section{Solving the Equatios of Motion}
\label{appendix:eom_sol}

In this appendix we give details on how the equations of motion are solved. It turns out to be convenient to write the resulting equations in terms of $A$ instead of the radial coordinate $z$. This change of coordinates stretches distances close to the boundary which eases the numerical UV analysis.
To implement this one introduces the new variable
\begin{equation}
q(A) = \frac{dz}{dA} \,e^{A} \,,
\label{eq: q definition}
\end{equation}
so that the metric reads
\begin{equation}
 ds^2 = q^2 dA^2 + e^{2A}\, \eta_{\mu\nu}dx^\mu dx^\nu \,.
\end{equation}
From now on all the background fields will be functions of $A$ and the differential equations we present are all with respect to $A$ being the independent variable.

\subsubsection*{Yang-Mills Equations of Motion}

If we set $x = 0$ we obtain the action of the IHQCD model and hence we are dealing with a pure Yang-Mills theory in the large $N_c$ limit. The equations of motion are then
\begin{align}
&\frac{dq}{dA} = \frac{1}{3} \left( 12 q - q^3 V_g (\Phi) \right) \, , \label{eq:q_YM_eom} \\
&\frac{d\Phi}{dA} = - \frac{\sqrt{3}}{2} \sqrt{12 - q^2 V_g(\Phi)} \label{eq:Phi_YM_eom}\,.
\end{align}
The solution of the pure Yang-Mills background is needed in our method for constructing the full V-QCD solution, as we explain below. 
We will therefore solve this coupled set of differential equations by shooting from the IR to the UV. 
The coordinate $A$ runs from $-\infty$ to $+\infty$, but naturally we will need to intriduce cutoffs for the numerical solution.
We set $A_{\mathrm{IR}} = -150$ and $A_{\mathrm{UVYM}} = 50$ as the lower and upper bound on $A$. In the IR the geometry ends in an IR singularity of the ``good'' kind according to the classification of~\cite{Gubser:2000nd}. At the singularity the warp factor $A$ and the dilaton $\Phi$ have the asymptotic forms~\cite{Jarvinen:2011qe}
\begin{align}
&A_{\mathrm{IR}} \left(z\right) =  - z^2   + \frac{1}{4} \log(6 z^2) - \frac{\log(\VgIRsymbol)}{2}+ \frac{23}{24}- \frac{173}{3456 z^2} +\mathcal{O}\left(\frac{1}{z^4}\right) , \\
&\Phi_{\mathrm{IR}} \left(z \right) = +\frac{3}{2} z^2  - \frac{23}{16}  - \log\left(\frac{\scsymbol}{\lambda_0}\right)- \frac{151}{2304 z^2}+\mathcal{O}\left(\frac{1}{z^4}\right) ,
\label{eq: YM fields IR asymptotics}
\end{align}
as $z \to \infty$. This allows us to determine $z_{\mathrm{IRYM}}$ such that $A_{\mathrm{IR}}\left(z_{\mathrm{IRYM}}\right) = -150$ and use its numerical value to compute $\Phi\left(A_{\mathrm{IR}}\right) = \Phi_{\mathrm{IR}} \left( z_{\mathrm{IRYM}}\right)$ and $q\left(A_{\mathrm{IR}}\right) = e^{A_{\mathrm{IR}}} / \frac{dA_{\mathrm{IR}}}{d z} $ and hence defining boundary conditions to solve equations~\eqref{eq:q_YM_eom} and~\eqref{eq:Phi_YM_eom}. Note that $\scsymbol$ and $\VgIRsymbol$ determine uniquely both the initial conditions as well as the evolution of the background fields in pure YM.

After we solve the YM equations we can determine $z\left(A\right)$ by using equation (\ref{eq: q definition}). We solve this numerically by taking $z(A_{\mathrm{IRYM}})$ = $z_{\mathrm{IRYM}}$ as initial condition and at the end we perform the shift $z\left(A\right) \to z\left(A\right) - z\left(A_{\mathrm{UVYM}} \right) $ in order to have the UV singularity at $z = 0$. 

We finish this section by specifying which algorithms we have used to find the numerical YM background. To compute $z_{\mathrm{IRYM}}$ we have used the root finding Van Wijngaarden-Dekker-Brent method described in~\cite{10.5555/1403886}. The differential equations~\eqref{eq:q_YM_eom} and~\eqref{eq:Phi_YM_eom} were solved using $\mathrm{integrate\_const}$ with a Runge-Kutta-Dormand-Prince stepper from the Boost $C\texttt{++}$ library.

\subsubsection*{Solving the equations of motion in the holographic model}

The solution of the equations of motion in V-QCD at zero temperature is done in four stages. This is because, as it turns out, the equations of motion are stiff in particular close to the IR singularity. Therefore for the numerical code to be stable, it is better to divide the range of $A$ into region where different kind of approximations can be used that reduce the stiffness problem. In the first two regions we are deep in the IR and the tachyon is decoupled from the other background fields and $q\left(A\right)$ so that $\Phi\left(A\right)$ obey equations (\ref{eq:q_YM_eom}) and (\ref{eq:Phi_YM_eom}). In these stages we therefore use the YM background constructed as explained above for the metric and for the dilaton. In the third stage the tachyon will couple to the other background fields until it reaches the UV where it decouples again starting the fourth and last stage of the solution of the EOMs.

As in the YM case we start by first specifying the IR boundary conditions. We first define $A_{\mathrm{IR}} = -150$, $A_{\mathrm{UVYM}} = 50$, $A_{\mathrm{UVc}} = 100$ and $A_{\mathrm{UVf}} = 1000$.
Again  we compute $z_{\mathrm{IRYM}}$ as in the YM case and we use it to compute $q\left(A_{\mathrm{IR}}\right)$ and $\Phi\left(A_{\mathrm{IR}}\right)$ as in YM and $\tau\left(A_{\mathrm{IR}}\right)$ using the tachyon IR asymptotics
\begin{equation}
\tau \sim \tau_0 \, z^{\tau_c}, \quad \tau_c = \frac{\left(12 - x W_0\right) \kIRsymbol a_2}{ 8\VgIRsymbol (a_2 - a_1)}\,,
\end{equation}
where $\tau_0$ is a parameter that is going to be fitted to the spectrum. 
The constants $\tau_{\mathrm{cut}} = 1000$ and $V_{\mathrm{f \, cut}} = 10^{-8}$ are defined and they mark the end of the first and second stages of the construction of the numerical background, respectively. The difference between the first and the second stage is that in the first stage, the tachyon is so large that nonlinear corrections $\sim 1/\tau^2$ can be ignored in the tachyon equation of motion. In the second stage we need to use the full tachyon equation of motion, while the tachyon remains decoupled (as signaled by the smallness of the tachyon potential, $V_f/V_g < V_{\mathrm{f \, cut}}$). 

We start the construction of the background by computing the YM profile of $q\left(A\right)$ and $\Phi\left(A\right)$ from $A_{\mathrm{IR}}$ to $A_{\mathrm{UVYM}}$ as explained above. If $\tau_{\mathrm{IR}} > \tau_{\mathrm{cut}}$ the tachyon profile will be given by the solution of the differential equation (linearized at large $\tau$)
\begin{equation}
	\frac{d\tau}{dA} =  \frac{2 q^2 V_{f0} \frac{d V_{\tau}}{d\tau} } {V_\tau \left(8 V_{f0} \ksymbol + 2 \ksymbol \frac{d V_{f0}}{d\Phi} \frac{d\Phi}{dA} + V_{f0}  \frac{d\ksymbol}{d\Phi} \frac{d\Phi}{dA} \right) }\,,
\end{equation}
until $\tau < \tau_{\mathrm{cut}}$, marking the end of the first stage. The value of $A$ such that this condition is met is called $A_{\mathrm{UV1}}$ The values of $q$ and $\Phi$ used are the ones given by YM.

In the second stage we solve the (full) tachyon differential equation
\begin{align}
&\frac{d^2\tau}{d A^2} =  \frac{q^2 \frac{V_\tau}{d\tau}}{ V_{\tau} \, \ksymbol} - 4 \frac{d \tau}{dA} + \frac{d \log q}{dA} \frac{d\tau}{dA} + \frac{\frac{V_\tau}{d\tau}}{V_{\tau}} {\left(\frac{d\tau}{dA}\right)}^2 - 4 \ksymbol \frac{{\left( \frac{d\tau}{dA}\right)}^3}{q^2} - \\ \notag
& - \frac{d \log V_{f0}}{d\Phi} \frac{d\tau}{dA} \frac{d\Phi}{dA} - \frac{d \log \ksymbol}{d\Phi} \frac{d\tau}{dA} \frac{d\Phi}{dA} - \ksymbol \frac{d \log V_{f0}}{d\Phi} {\left(\frac{d\tau}{dA}\right)}^3 \frac{\frac{d\Phi}{dA}}{q^2} - \frac{d \ksymbol}{d\Phi} \frac{d\Phi}{dA} \frac{{\left(\frac{d\tau}{dA}\right)}^3}{2 q^2}\,,
\label{eq: tau YM 2}
\end{align}
from the value of $A_{\mathrm{UV1}}$ to $A_{\mathrm{UVYM}}$. Using the profiles of $q$, $\Phi$ and $\tau$ we compute $A_{\mathrm{UV2}}$ such that $A_{\mathrm{UV1}} < A_{\mathrm{UV2}} < A_{\mathrm{UVYM}}$ and $V_{f}(\Phi,\tau) = V_{f0}\left(\Phi\right) V_\tau(\tau) = V_{\mathrm{f \, cut}} V_g \left(\Phi\right)$ are satisfied. This condition marks when the tachyon starts to couple with $q$ and $\Phi$ from the IR to the UV.

In the stage where the tachyon is coupled the dynamics of the background fields  obeys
\begingroup
\allowdisplaybreaks
\begin{align}
& \frac{dq}{dA} = \frac{4}{9} q {\left(\frac{d\Phi}{dA}\right)}^2 + x \, q V_f  \ksymbol \frac{{\left(\frac{d\tau}{dA}\right)}^2}{6\sqrt{1+ \ksymbol {\left(\frac{\frac{d\tau}{dA}}{q}\right)}^2}} \,,\\
& \frac{d^2 \Phi}{d^2A} = - \frac{3}{8} q^2 \frac{d V_g}{d\Phi} + \frac{9}{\frac{d\Phi}{dA}} + \frac{3}{4} \frac{q^2}{\frac{d\Phi}{dA}} \left( \frac{x V_f }{\sqrt{1+ k \frac{{\left(\frac{d\tau}{dA}\right)}^2}{q^2}}} - V_g  \right) - \\ \notag
& - 5 \frac{d\Phi}{d A} + x V_f  k  \frac{{\left(\frac{d\tau}{dA}\right)}^2 \frac{d\Phi}{dA}}{6 \sqrt{1+ \ksymbol {\left(\frac{\frac{d\tau}{dA}}{q}\right)}^2 }} + \frac{4}{9} {\left(\frac{d\Phi}{dA}\right)}^3 + \\ \notag
& + \frac{3}{8} x q^2 \frac{\frac{d V_f}{d\Phi}}{\sqrt{1+ \ksymbol {\left(\frac{\frac{d\tau}{dA}}{q}\right)}^2}} + \frac{3}{8} x \ksymbol {\left( \frac{d\tau}{dA}\right)}^2 \frac{\frac{dV_f}{d\Phi}}{\sqrt{1+ \ksymbol {\left(\frac{\frac{d\tau}{dA}}{q}\right)}^2}} + \\ \notag
& + \frac{3}{16} x V_f {\left( \frac{d\tau}{dA}\right)}^2 \frac{\frac{d\ksymbol}{d \Phi}}{\sqrt{1+ \ksymbol {\left(\frac{\frac{d\tau}{dA}}{q}\right)}^2}} \,,\\ \notag
& \frac{d^2 \tau}{d A^2} =  \frac{q^2 \frac{d V_\tau}{d\tau}}{V_{\tau} \, \ksymbol} - 4 \frac{d\tau}{d A} + \frac{\frac{dV_{\tau}}{d\tau}}{V_{\tau}} {\left(\frac{d\tau}{dA}\right)}^2 - 4\frac{\ksymbol}{q^2} {\left(\frac{d\tau}{dA}\right)}^3 + x \ksymbol V_f \left(\Phi, \tau \right) \frac{{\left(\frac{d\tau}{dA}\right)}^3}{6 \sqrt{1+ \ksymbol {\left(\frac{\frac{d\tau}{dA}}{q}\right)}^2}} - \\
& - \frac{d \log(V_{f0})}{dA} \frac{d\Phi}{dA} \frac{d\tau}{dA} - \ksymbol  \frac{d\log V_{f0}}{dA} \frac{d\Phi}{dA}\frac{{\left(\frac{d\tau}{d A}\right)}^3}{q^2} + \frac{4}{9} {\left(\frac{d\Phi}{dA}\right)}^2 \frac{d\tau}{dA} - \frac{d \log(\ksymbol)}{d\Phi} \frac{d\Phi}{dA} \frac{d\tau}{dA} - \\ \notag
& - \frac{1}{2} \frac{d \ksymbol}{d\Phi} \frac{d\Phi}{dA} \frac{{\left( \frac{d\tau}{dA}\right)}^3}{q^2}\,.
\end{align}
\endgroup
This is a stiff system of differential equations and for this reason we had used the function integrate\_adaptive with the $\mathrm{rosenbrock4}$ stepper from the Boost $C\texttt{++}$ library. This system is solved from $A_{\mathrm{UV2}}$ to $A_{\mathrm{UVc}}$. The value $A_{\mathrm{UVc}} = 50$ was chosen such that at this point we are in the UV and the tachyon decouples again from $q$ and $\Phi$: this is guaranteed as the tachyon is suppressed exponentially in the UV ($\tau \sim m_q e^{-A}$).

The UV equations of motion are
\begin{align} 
&\frac{dq}{dA} =  \frac{4}{9} q {\left(\frac{\frac{d\lambda}{dA}}{\lambda}\right)}^2 \,,\\
&\frac{d^2\lambda}{dA^2} = -\frac{3}{8}  {\left(q \lambda \right)}^2 \frac{d V_g}{d\lambda} + 9 \frac{\lambda^2}{\frac{d\lambda}{dA}} + \frac{3}{4} \frac{{\left(q \lambda \right)}^2}{\frac{d\lambda}{dA}} \left( x \, V_{f0} - V_g  \right) - 5 \frac{d\lambda}{dA}  + \\
& + \frac{d \log q}{dA} \frac{d\lambda}{d A} + \frac{{\left( \frac{d\lambda}{dA}\right)}^2}{\lambda} + \frac{3}{8} \, x \, {\left( q \lambda \right)}^2 \frac{d V_{f0}}{d\lambda} \,,\\ \notag
& \frac{d^2 \tau_n}{dA^2} = 3 \tau_n  +  \frac{e^A q^2 \frac{V_\tau}{d\tau} }{V_\tau \ksymbol} + \tau_n \frac{d \log V_{f0}}{d\lambda} \frac{d\lambda}{dA} + \tau_n \frac{d\lambda}{dA} \frac{d \log \ksymbol}{d\lambda} -\frac{d\log q}{d A}  \tau_n  - \\
& - 2 \frac{d\tau_n}{d A} - \frac{d \log V_{f0}}{d\lambda} \frac{d\lambda}{dA} \frac{d\tau_n}{dA}  - \frac{d\lambda}{dA} \frac{d\tau_n}{dA} \frac{d \log \ksymbol}{d\lambda} + \frac{d\log q}{d A}  \frac{d \tau_n}{dA} \, ,
\end{align}
where $\tau_n = e^A \tau$. This system is solved from $A_{\mathrm{UVc}}$ to $A_{\mathrm{UVf}}$. With the profile of $\tau_n$ and $\lambda$ in this region we can compute an estimate the quark mass $m_q$ through the UV asymptotic formula
\begin{align}
  m_{q\,\mathrm{est}}(A) &= \frac{1}{\ell_{\mathrm{UV}}} \tau_n\left(A\right)e^{- \tau_{corr}\left(\lambda\left(A\right)\right)}  \,,&\\
\tau_{\mathrm{corr}}(\lambda) &= \frac{\left(-88 + 16 x + 27 \scsymbol \, \kUpsymbol\right) \log\left( \frac{24 \pi^2}{\left(11 - 2 x\right) \lambda}\right)}{12 x - 66}\,,
\end{align}
where $\ell_\mathrm{UV}$ is the UV AdS radius.
We then obtain the final estimate for the quark mass by evaluating this estimate at two large values, i.e. at $A=A_\mathrm{UVf}$ and at $A=A_\mathrm{UVf}-10$ and linearly extrapolating the result to the UV (i.e. $\lambda=0$) on the ($\lambda,m_{q\,\mathrm{est}}$) -plane.
As in the YM case we can determine $z\left(A\right)$ by using equation (\ref{eq: q definition}) and perform the shift $z\left(A\right) \to z\left(A\right) - z\left(A_{\mathrm{UVYM}} \right) $ in order to have the UV singularity at $z = 0$.

\section{EOM and couplings of the $U(1)$ gauge field}
\label{appendix:b}

In this appendix we start deriving the action of the vector meson sector since the non-normalisable solution of the corresponding equations of motion is dual to the external photon in the boundary. Then, by linearising the action we find the coupling of the $U(1)$ gauge field to the graviton of the bulk theory. All of this is made in the Einstein frame. Since we will be dealing later with calculations on the string frame we will explain how to translate these results to the string frame. Finally we generalise the coupling to the graviton to any spin $J$ field in the graviton's Regge trajectory.

\subsection*{The action of the vector $U(1)$ gauge field}
As mentioned previously, for the QCD vacuum we set $A^R_a = 0 = A^L_a$.
When we turn on the gauge fields, the flavour action becomes
\begin{align}
&S_f = - \frac{M^3 N_c}{2}  \int d^5 x \,V_f ( \lambda, \tau) \bold{Tr} \left[ \sqrt{-\mathrm{det} \left((g_{\mathrm{eff.}})_{ab} + T^{L}_{ab}\right)} + \sqrt{-\mathrm{det} \left((g_{\mathrm{eff.}})_{ab} + T^{R}_{ab}\right)} \right] \, , \notag \\
&(g_{\mathrm{eff.}})_{ab} = g_{ab} +  \ksymbol\left(\lambda\right) \partial_a \tau \partial_b \tau \, , \notag \\
&T^{L}_{ab} = w\left(\lambda\right) F^{L}_{ab} + \ksymbol\left(\lambda\right)D_{(a}\tau D_{b)}\tau -   \ksymbol\left(\lambda\right) \partial_a \tau \partial_b \tau \, , \\
&T^{R}_{ab} = w\left(\lambda\right) F^{R}_{ab} + \ksymbol\left(\lambda\right)D_{(a}\tau D_{b)}\tau - \ksymbol\left(\lambda\right) \partial_a \tau \partial_b \tau \, .\notag
\end{align}
Here we remember that $\bold{Tr}$ is a trace over the flavour indices while the determinant is computed with respect to the space-time indices. Writing the determinants inside the square roots as 
\begin{align}
\mathrm{det}\left(g_{\mathrm{eff.}} + T^{L/R} \right) = \mathrm{det} \left(g_{\mathrm{eff.}} \right)  \mathrm{det} \left( 1 + g_{\mathrm{eff.}}^{-1} T^{L/R} \right) \, ,
\end{align}
defining $X^{L/R} = g_\mathrm{eff.}^{-1} T^{L/R}$ and using the identity
\begin{align}
\ln \text{det} \left(1+X^{L/R}\right) = \text{tr} \ln\left(1 + X^{L/R}\right) = \text{tr} X^{L/R} - \frac{1}{2} \text{tr} {\left(X^{L/R}\right)}^2 + \cdots \, ,
\end{align}
it follows that
\begin{align}
&\sqrt{\text{det}  \left(1+X^{L/R}\right)} = 1 + \frac{1}{2} \text{tr} \, X^{L/R} - \frac{1}{4} \text{tr} \, {\left(X^{L/R}\right)}^2 + \cdots \, , \\
& \text{tr} \, X^{L/R} = \sum_{a,b}  \left( g^{-1}_{\mathrm{eff.}} \right)_{ab} T^{L/R}_{ab} \, , \\
& \text{tr} \, {\left(X^{L/R}\right)}^2 = \sum_{a,b,c,d}  \left( g^{-1}_{\mathrm{eff.}} \right)_{ac} T^{L/R}_{cb} \left( g^{-1}_{\mathrm{eff.}} \right)_{bd} T^{L/R}_{da} \, .
\end{align}
The expressions for the matrix elements of  $g^{-1}_{\mathrm{eff.}}$ and $T^{L/R}$ are
\begin{align}
&\left(g^{-1}_{\mathrm{eff.}}\right)_{zz} = \frac{e^{-2A}}{G^2} \quad , \quad  \left(g^{-1}_{\mathrm{eff.}} \right)_{\mu \nu} = \eta^{\mu \nu} e^{-2A} \, , \\
&\left(g^{-1}_\mathrm{eff.}\right)_{z\mu} = \left(g^{-1}_\mathrm{eff.}\right)_{\mu z} = 0\,, \\
& T^{L/R}_{zz} = 0 \, , \\
& T^{L/R}_{z \mu} = - T^{L/R}_{\mu z} =  w\left(\lambda \right) \left(  \pm \partial_z A_\mu + \partial_z V_\mu \right) \, , \\
& T^{L/R}_{\mu \nu} = 4 \ksymbol\left(\lambda\right) \tau^2 A_\mu A_\nu + w\left(\lambda \right) \left( \pm A_{\mu \nu} + V_{\mu \nu} \right) \, ,
\end{align}
where the vector and axial gauge fields $V_a$ and $A_a$ are linear combinations of the left and right gauge fields
\begin{equation}
V_a = \frac{A^L_a + A^R_a}{2}\, , \qquad  \, A_a = \frac{A^L_a - A^R_a}{2}\,,
\end{equation}
in the gauge $V_z = 0 = A_z$, $V_{\mu \nu} = \partial_\mu V_\nu  - \partial_\nu V_\mu$ and $A_{\mu \nu} = \partial_\mu A_\nu  - \partial_\nu A_\mu$.

From these identities and after some 
calculations one gets
\begin{align}
&S_f = - \frac{1}{2} M^3 N_c \int d^5 x\, V_f (\lambda, \tau) \sqrt{- g_\mathrm{eff.}}\, \bold{Tr} \bigg[ 2 + 4 e^{-2A} \ksymbol\left( \lambda \right) \tau^2 A_\mu A^\mu + \\
&  + \frac{{w\left(\lambda\right)}^2 e^{-4A}}{G^2} \left( \partial_z A_\mu \partial_z A^\mu +  \partial_z V_\mu \partial_z V^\mu \right) + \frac{1}{2} e^{-4A} {w\left(\lambda\right)}^2 \left( V_{\mu \nu}V^{\mu \nu} + A_{\mu \nu} A^{\mu \nu} \right) \bigg] \, .\notag 
\end{align}
The first term is the background term, while the the two other terms can be packed into the actions
\begin{align}
\label{eq:VM_action}
S_V =& - \frac{1}{2} M^3 N_c \bold{Tr} \int d^5 x \,V_{f} ( \lambda, \tau ) \,{w(\lambda)}^2 G^{-1} e^{A} \left( \partial_z V_\mu \partial_z V^\mu + \frac{1}{2} G^2 V_{\mu \nu} V^{\mu \nu} \right), \\
 S_A =& - \frac{1}{2} M^3 N_c \bold{Tr} \int d^5 x \,V_{f} ( \lambda, \tau ) \,{w(\lambda)}^2 G^{-1} e^{A} \bigg( \partial_z A_\mu \partial_z A^\mu + \notag \\
& \qquad\qquad \qquad \qquad \qquad \qquad 
+ \frac{1}{2} G^2 A_{\mu \nu} A^{\mu \nu} + 4 e^{2A} \frac{\ksymbol (\lambda) G^2 \tau^2}{{w(\lambda)}^2} A_\mu A^\mu \bigg)\,,
\end{align}
which are the actions for the Non-singlet Vector and Axial-Vector mesons presented in~\cite{Arean:2013tja}.
The equation (\ref{eq:VM_action}) can still be rewritten as
\begin{equation}
S = - \frac{1}{4} M^3 N_c N_f \int d^5 x \sqrt{-g} \,V_f\, w^2 G F_{ab}F^{ab},
\label{eq:VM_new_action}
\end{equation}
where here the notation $F_{ab}F^{ab}$ means
\begin{equation}
F_{ab}F^{ab} =  \sum_{a,b,c,d}F_{ab} \left(g^{-1}_{\mathrm{eff.}}\right)_{ac} \left(g^{-1}_{\mathrm{eff.}}\right)_{bd} F_{cd} \,.
\end{equation}

\subsection*{From Einstein frame to string frame}

To compute the action of the $U(1)$ gauge field in the Einstein frame we approximated the square roots of the determinants present in $S_f$ as
\begin{align}
\sqrt{-\det \left( g_{\mathrm{eff.}} + T^{L/R} \right)} = \sqrt{-g_{\mathrm{eff.}}} \left[ 1 + \frac{1}{2} \text{tr} X - \frac{1}{4} \text{tr} X^2 + \frac{1}{8} {\left( \text{tr} X \right)}^2 + \dots \right],
\end{align}
with $X = g^{-1}_{\mathrm{eff.}} T^{L/R}$ for some tensor $T^{L/R}$.
The string frame warp factor $A_s$ is related to the Einstein frame warp factor $A$ through $A_s = A + \frac{2}{3} \Phi$. From this it follows that $\sqrt{-g_{\mathrm{eff.}}} = e^{- 10 \Phi / 3} \sqrt{-g_{\mathrm{eff.}\, s}}$ and the matrix $X$ in the previous equation can be written as $X = g_s^{-1} T^{L/R}_s$ with $T^{L/R}_s = e^{4 \Phi / 3} T^{L/R}_E$. This implies that
\begin{align}
\sqrt{-\det \left( g_{\mathrm{eff.}} + T^{L/R} \right)} = e^{- 10 \Phi / 3} \sqrt{-\det \left( g_{\mathrm{eff.}\, s} + T^{L/R}_s \right)} \, ,
\end{align}
i.e.  $S_f$ in the string frame is obtained by simply substituting the metric by the string frame metric $g_s$, substituting $\ksymbol(\lambda)$ and $w(\lambda)$ by $\ksymbol _s(\lambda) = e^{4 \Phi / 3} \ksymbol(\lambda) $ and $w_s (\lambda) = e^{4 \Phi / 3} w(\lambda) $ and multiplying $V_f$ by the factor $e^{-10 \Phi / 3}$. The derivation of $S_V$ in the string frame will be formally the same and hence the results equal to the Einstein frame ones but with $w_s$, $\ksymbol _s$ and $e^{-10 \Phi / 3} V_f$ in place of $w$, $\ksymbol$ and $V_f$ respectively. The action (\ref{eq:VM_new_action}) in the string frame takes the form
\begin{equation}
S = - \frac{1}{4} M^3 N_c N_f \int d^5 x \sqrt{-g_s} e^{-\frac{10}{3} \Phi}\, V_f\, w_s^2 G F_{ab}F^{ab}  \, .
\label{eq:VM_new_action_string_frame_2}
\end{equation}

\subsection*{Couplings with the spin $J$ fields}

We will now determine the gravitational coupling between the $U(1)$ gauge field and the spin $J$ fields in the graviton's Regge trajectory. We first compute the coupling with the graviton and generalise to any even spin $J$ field. All of this is done in the Einstein frame. To find the coupling in the string frame we just substitute the functions $w$, $\ksymbol$ and $V_f$ by $w_s$, $\ksymbol _s$ and $e^{-10 \Phi / 3} V_f$ respectively, as discussed previously. 

Again, we start by writing the square roots of the determinants as
\begin{equation}
	\sqrt{-\det g_{\mathrm{eff.}}} \left[ 1 + \frac{1}{2} \text{tr} \left( g^{-1}_{\mathrm{eff.}} T^{L/R}\right) - \frac{1}{4} \text{tr} \left( g^{-1}_{\mathrm{eff.}} T^{L/R} g^{-1}_{\mathrm{eff.}} T^{L/R} \right) + \dots \right].
\label{eq:det_geff_expanded}
\end{equation}
The coupling with the graviton is found by linearising equation (\ref{eq:det_geff_expanded}) around the background metric, i.e. $g_{ab} = \bar{g}_{ab} + h_{ab}$. 
To study the graviton Regge trajectory in our background we need to decompose the metric in $SO(1,3)$ irreducible representations. 
We will be only interested in the graviton $\mathrm{TT}$ components $h_{\alpha \beta}$, satisfying $\partial^\alpha h_{\alpha \beta} = 0$ and $h^\alpha_\alpha = 0$, and also set $h_{z \alpha} = h_{\alpha z} = h_{zz} = 0$.

 For our purposes we can ignore the perturbation of $\sqrt{-\det g_{\mathrm{eff.}}}$ because it involves only a term proportional to $h = h^a_a = 0$. We will also neglect the terms involving the axial vector mesons $A_{\mu}$ since we are only interested in the coupling of $V_\mu$ with the graviton for now. We wish then to compute $\delta \text{tr} \left( g^{-1}_{\mathrm{eff.}} T^{L/R}\right)$ and $\delta \text{tr} \left( g^{-1}_{\mathrm{eff.}} T^{L/R} g^{-1}_{\mathrm{eff.}} T^{L/R} \right)$, where by $\delta$ we mean a perturbation relative to the background metric. Using the identity $\delta g^{ab}_{\mathrm{eff.}} = - g_{\mathrm{eff.}}^{am} g_{\mathrm{eff.}}^{bn} h_{mn}$ one can show that
\begin{align}
&\delta \text{tr} \left( g^{-1}_{\mathrm{eff.}} T^{L/R}\right) = -  g_{\mathrm{eff.}}^{am} g_{\mathrm{eff.}}^{bn} h_{mn} T^{L/R}_{ab}\, , \\
&\delta \text{tr} \left( g^{-1}_{\mathrm{eff.}} T^{L/R} g^{-1}_{\mathrm{eff.}} T^{L/R} \right) = - 2 g^{am}_{\mathrm{eff.}} g^{bn}_{\mathrm{eff.}} g^{cd}_{\mathrm{eff.}} T^{L/R}_{bc} T^{L/R}_{da} h_{mn}\, .
\end{align}
Using the expressions for the matrix elements of $g^{-1}_{\mathrm{eff.}}$ and $T^{L/R}$ we get
\begin{align}
&\delta \text{tr} \left( g^{-1}_{\mathrm{eff.}} T^{L/R}\right) = 0 \, , \\
& \delta \text{tr} \left( g^{-1}_{\mathrm{eff.}} T^{L/R} g^{-1}_{\mathrm{eff.}} T^{L/R} \right)  = 2 e^{-6A} {w\left( \lambda \right)}^2 h^{\mu \nu} \left( V_{\mu \sigma} \eta^{\sigma \rho} V_{\nu \rho} + \frac{1}{G^2} \partial_z V_\mu \partial_z V_\nu \right) .
\end{align}
Hence the coupling between the vector $U(1)$ gauge field and the graviton is given by
\begin{align}
\frac{M^3 N_c N_f}{2}  \int d^5 x \sqrt{-\det g_{\mathrm{eff.}}} V_{f} (\lambda, \tau)\, e^{-6 A} {w(\lambda)}^2 h^{\mu \nu} \left( V_{\mu \sigma} \eta^{\sigma \rho} V_{\nu \rho} + \frac{1}{G^2} \partial_z V_\mu \partial_z V_\nu \right),
\end{align}
or simply
\begin{align}
\frac{M^3 N_c N_f}{2} \int d^5 x \sqrt{-g}\, G \,V_{f} (\lambda, \tau )  {w\left(\lambda\right)}^2 g^{c m} g^{d n} g_\mathrm{eff.}^{ab} F^V_{a c} F^V_{b d} h_{m n} \, .
\end{align}
In the string frame this coupling takes the form
\begin{align}
\frac{M^3 N_c N_f}{2} \int d^5 x \sqrt{-g_{s}} \,G \,e^{-\frac{10}{3}\Phi} V_{f} (\lambda, \tau ) \, {w_s(\lambda)}^2 g^{c m}_s g^{d n}_s g_{\mathrm{eff.}\,s}^{ab} F^V_{a c} F^V_{b d} h_{m n}^s \, .
\end{align}
We now generalise this coupling to the case of an interaction between the gauge field and a symmetric, transverse and traceless spin $J$ field, $h_{a_1 \cdots a_J}$. The pomeron trajectory includes such higher spin fields of even $J$. Again there are several possibilities, but we shall focus on the simplest extension of the graviton coupling considered above.
For a spin $J$ field we take the coupling
\begin{align}
k_J \int d^5x \sqrt{-g_s} \,G \,e^{-\frac{10}{3} \Phi} V_f ( \lambda, \tau )\, {w_s(\lambda )}^2  g_{\mathrm{eff.}\,s}^{ab} F^V_{a c} \nabla_{a_1} \dots \nabla_{a_{J-2}} F^V_{b d} h^{c d a_1 \dots a_{J-2}}\,.
\label{eq:spin_J_graviton_coupling}
\end{align}
We note that the transverse condition of the spin $J$ field $h_{a_1 \cdots a_J}$ guarantees that this term is unique up to dilaton and tachyon derivatives.

We now consider the coupling between the external photon states with the bulk spin $J$ fields dual to the spin $J$ twist two operators made of quark bilinears. To determine the coupling to any spin $J$ in this trajectory we could proceed analogously with the case of the graviton's Regge trajectory. In the case of the coupling with the meson trajectory, one could first determine the coupling between the non-normalizable mode dual to the photon with the the $\rho$ meson states and generalise the result to higher spin $J$ fields. To do this we attempted to expand the DBI action to cubic order in the fluctuations and keep only the terms with three vector gauge fields $V_a V_b V_c$. We start by writing
\begin{align}
\sqrt{\mathrm{det} (g_{\mathrm{eff.}} + T^{L/R} )} = \sqrt{\mathrm{det} g_{\mathrm{eff.}}} \exp{ \left[ \frac{1}{2} \mathrm{Tr} \log\left(1 + X^{L/R}\right) \right]} \,.
\end{align}
If we use the power series expansion of the exponential and of the logarithmic function we get
\begin{align}
\sqrt{-\mathrm{det} g_{\mathrm{eff.}}} \left[1- \frac{1}{4} \mathrm{Tr} \left( X^{L/R}\right)^2 + \frac{1}{6}\mathrm{Tr} \left( X^{L/R}\right)^3  \right] \, ,
\end{align}
where we have dropped terms involving products of $\mathrm{Tr} X^{L/R}$ because they contribute only to axial gauge fields $A_a$. The quadratic term leads to the action of the quadratic fluctuations of the vector gauge field and hence the coupling $V_a V_b V_c$ if exists must be contained on $\mathrm{Tr} \left( X^{L/R}\right)^3$. That is, we evaluated the expression
\begin{align}
\mathrm{Tr} \left( X^{L/R}\right)^3 = \sum_{i, k, j, l, m, n}  \left( g^{-1}_{\mathrm{eff.}} \right)_{ik}  \left( g^{-1}_{\mathrm{eff.}} \right)_{ln}  \left( g^{-1}_{\mathrm{eff.}} \right)_{jm} T^{L/R}_{kj}  T^{L/R}_{ml}  T^{L/R}_{ni} \, ,
\end{align}
which contains no coupling of the form  $V_a V_b V_c$.

Another approach is to find the coupling between the vector gauge field with the bulk field dual to the $f_2$ meson and extrapolate to all the other spin $J$ fields in the meson trajectory. In~\cite{Katz:2005ir} the tensor meson $f_2$ state is the first Kaluza-Klein mode of a bulk spin-2 $h_{ab}$ field that has the same equation of motion as the graviton in $\mathrm{AdS}_5$. The coupling of $f_2$ to the photon is also the same as the one between a graviton and a bulk gauge field in $AdS_5$. The geometry is basically $AdS_5$ with a wall whose position is fixed by the mass of the $\rho$ meson. After this they are able to predict not only the mass of $f_2$ but also the decay width $\Gamma(f_2 \to \gamma \gamma)$. $f_2$ also has the same quantum numbers of the tensor glueballs  $J^{PC} = 2^{++}$ which are the normalisable modes associated with the graviton's equation of motion.  For these reasons, in this work, we will assume that the coupling of the $U(1)$ gauge field with the $f_2$ meson is the same as the coupling with the graviton and hence, in general, the coupling of any bulk spin $J$ field in the meson trajectory is also given by equation~\eqref{eq:spin_J_graviton_coupling}.

\section{$pp$ scattering}
\label{sec:pp_scattering}

In this appendix we will present the computation for the total cross-section of $pp$ scattering. The steps of the computation are the same as in the case of $\gamma p$ show in the main text.

The scattering amplitude for spin $J$ exchange between two incoming scalar fields $\Upsilon^{(1)} \sim e^{i k_1 \cdot x}$ and $\Upsilon^{(2)} \sim e^{i k_2 \cdot x}$ is
\begin{equation}
\mathcal{A}_J = {\left(\bar{k}_J\right)}^2 \int d^5x d^5\bar{x} \sqrt{-g} \sqrt{-\bar{g}} \,e^{- \Phi -\bar{\Phi}} \left(\Upsilon_1 \partial_{-}^J \Upsilon_3\right) \Pi^{-\cdots-, + \cdots +} \left(x, \bar{x}\right) \left(\bar{\Upsilon}_2 \bar{\partial}_{+}^J \bar{\Upsilon}_4\right)  ,
\end{equation}
where it was taken into account that the kinematics~\eqref{eq:pp_kinematics} implies that in the Regge limit the component $\Pi^{-,\cdots-, +\cdots+}$ dominates. Lowering the indices of the spin $J$ propagator, making the change of variable $w = x - \bar{x}$ and using the identity
\begin{equation}
\int d^2 l_\perp e^{- i q_\perp \cdot l_\perp}\int \frac{dw^+ dw^-}{2} \Pi_{+\cdots+, - \cdots -} \left(x, \bar{x}\right) = - \frac{i}{\left(-2\right)^J} e^{\left(J-1\right)\left(A+\bar{A}\right)} G_J \left(z, \bar{z}, t\right),
\end{equation}
after some algebra the scattering amplitude can be rewritten as
\begin{align}
\mathcal{A}_J = - i V \, \frac{\bar{k}_J^2}{2^J} \,s^J \int dz d\bar{z} \,e^{4\left(A+\bar{A}\right)} e^{-J\left(A+\bar{A}\right)} e^{-\Phi - \bar{\Phi}} {|\upsilon_1|}^2 {|\upsilon_2|}^2  G_J (z, \bar{z}, t)\,.
\end{align}

As in the $\gamma^*p$ case, in order to get the total amplitude we need to sum over the spin $J$ fields with $J \geq J_{min}$, where $J_{min}$ is the minimal spin in the corresponding Regge trajectory. 
Then we can apply a Sommerfeld-Watson transform
\begin{align}
\frac{1}{2} \sum_{J \geq J_{min}} \left(s^J + {\left(-s\right)}^J\right) \frac{\mathcal{A}_J}{s^J} = - \frac{\pi}{2} \int \frac{d J}{2 \pi i} \frac{s^J + \left(-s\right)^J}{\sin \pi J} \frac{\mathcal{A}_J}{s^J}\,,
\end{align}
where we are assuming the analytic continuation of the scattering amplitude $\mathcal{A}_J$ to the complex $J$-plane. Deforming the $J$-plane integral and catching all the poles $J = j_n(t)$ defined by $t_n(J) = t$ we get
\begin{align}
\mathcal{A} = \frac{\pi}{2} \sum_n \frac{\bar{k}_{j_n}^2}{2^{j_n}} \,s^{j_n} \left[i + \cot \left( \frac{\pi j_n}{2} \right) \right] \frac{d j_n}{dt} {\left| \int dz \,e^{-\left(j_n - 4\right) A} {|\upsilon_1|}^2 e^{-\Phi} e^B \psi_n(z) \right|}^2\,.
\end{align}
In the scattering domain of $t < 0$ these poles are in the real axis for $J < J_{min}$. This procedure yields equations (\ref{eq:sigma_pp_hol_gens}) and (\ref{eq:pp_couplings_def}).

\section{$\gamma^{*} \gamma$ processes}
\label{sec:gg_scattering}

In this section we derive the holographic expressions for $F_2^\gamma$ and $\sigma(\gamma \gamma \rightarrow X)$ in the context of Holographic QCD in the Veneziano limit. We will consider the photon structure function $F^{\gamma}_2$ and the total cross-section $\sigma(\gamma \gamma \to X)$. Like in the case of the proton structure function $F_2^p$ the photon strucutre function $F_2^\gamma$ is related to the transverse and longitudinal total cross-sections of $\gamma^{*}\gamma$ scattering by
\begin{equation}
F_2^\gamma = \frac{Q^2}{4 \pi^2 \alpha} \left( \sigma_T^{\gamma^* \gamma} + \sigma_L^{\gamma^* \gamma} \right).
\end{equation}

The calculation of the forward scattering amplitude for $\gamma^{*}\gamma$ is the same as in $\gamma^{*} p$ scattering except that the external state in the Witten diagram of figure~\ref{fig:Witten_diagrams} is an on-shell photon. This means that the definition of $\mathrm{Im}^{\gamma \gamma}_n$ should be proportional to $k^2_{j_n}$ instead of $k_{j_n}\bar{k}_{j_n}$ and the integral appearing in it should be
\begin{equation}
\int d\bar{u} \,e^{-\left(j_n - 2 \right) \bar{A} } e^{-\frac{10}{3} \bar{\Phi} }\, \bar{V}_{f} \, \bar{w}_s^2  \, e^{\bar{B}} \psi_n(\bar{u}) \, .
\end{equation}
Then, the holographic expressions for $F_2^\gamma$ is
\begin{align}
&F^{\gamma}_2\left(x, Q^2\right) = \sum_n \frac{{\mathrm{Im}} g^{\gamma \gamma}_n}{4 \pi^2 \alpha} \, Q^{2 j_n} x^{1-j_n}  \int du \,e^{-\left(j_n - 2 \right)A } e^{-\frac{10}{3} \Phi} \,V_f \,w_s^2  \left(f_Q^2 +\frac{\partial_u{f_Q}^2}{Q^2} \right) e^B \psi_n(u) \, \label{eq:hol_f2_gamma} ,
\end{align}
while the holographic expression for $\sigma\left(\gamma \gamma \to X\right)$ is (\ref{eq:sigma_gg_hol_gens}) and the definition of $g^{\gamma \gamma}_n$ is given by~\eqref{eq:gg_couplings_def}.

\section{Holographic structure functions} \label{appendix:pgammastr} 

The forward scattering amplitude for $\gamma^{*}p$ scattering is given by equation (\ref{eq:gp_forward_scattering_amp}). That equation was obtained summing the contributions of the transverse and longitudinal polarisations of the off-shell photon. In particular the term with $f_Q^2$ is the contribution from transverse polarisations while the term $\dot{f}_Q^2$ is the contribution for the longitudinal polarisation.

The proton structure functions $F^p_2$ and $F^p_L$ are related to the transverse and longitudinal total cross-sections of the process $\gamma^{*} p$ by
\begin{align}
&F_2\left(x, Q^2\right) = \frac{Q^2}{4 \pi^2 \alpha} \left(\sigma_T^{\gamma^* p} + \sigma_L^{\gamma^* p} \right) , \\
&F_L\left(x, Q^2\right) = \frac{Q^2}{4 \pi^2 \alpha} \sigma_L^{\gamma^* p} \,.
\end{align}
Using the optical theorem and the last relations one finds the contribution of the holographic expressions to the structure functions are
\begin{align}
&F_2\left(x, Q^2\right) = \sum_n \frac{{\mathrm{Im}} g^{\gamma p}_n}{4 \pi^2 \alpha} \, Q^{2 j_n} x^{1-j_n}  \int du\, e^{-\left(j_n - 2 \right)A } e^{-\frac{10}{3} \Phi} \,V_f\, w_s^2  \left(f_Q^2 +\frac{\partial_u{f_Q}^2}{Q^2} \right)  e^B \psi_n(u) \, \label{eq:hol_f2} , \\
&F_L\left(x, Q^2\right) = \sum_n \frac{{\mathrm{Im}} g^{\gamma p}_n}{4 \pi^2 \alpha} \, Q^{2 j_n} x^{1-j_n}  \int du \,e^{-\left(j_n - 2 \right)A } e^{-\frac{10}{3} \Phi} \,V_f \,w_s^2 {\partial_u{f_Q}^2}{Q^2} \, e^B \psi_n(u) \, \label{eq:hol_fl} ,
\end{align}
where the definition of $g^{\gamma p}_n$ is the one of equation (\ref{eq:gp_couplings_def}). The function $B$ 
will depend on whether the spin $J$ fields belong to the pomeron or meson trajectory.

The structure function $F^p_2$ is, as expected, related to the total cross-section $\sigma(\gamma p \rightarrow X)$ through
\begin{equation}
\sigma(\gamma p \rightarrow X) = 4 \pi^2 \alpha \lim_{Q^2 \rightarrow 0} \frac{F_2\left(x, Q^2\right)}{Q^2} \,.
\end{equation}

\normalem
\bibliographystyle{JHEP}
\bibliography{bib/HVQCD.bib}

\end{document}